\documentclass[twocolumn,trackchanges,twocolappendix]{aastex631}
\usepackage{amsmath,bm}

\received{}
\revised{}
\accepted{}

\shorttitle{injection and reacceleration of cosmic-ray protons in galaxy clusters}
\shortauthors{Nishiwaki and Asano}

\begin{document}

\title{Low injection rate of cosmic-ray protons in the turbulent reacceleration model of radio halos in galaxy clusters}

\author[0000-0003-2370-0475]{Kosuke Nishiwaki}
\affiliation{Institute for Cosmic Ray Research, The University of Tokyo, 5-1-5 Kashiwanoha, Kashiwa, Chiba 277-8582, Japan}
\affiliation{Department of Physics, Graduate School of Science, Tokyo Metropolitan University, 1-1 Minami-Osawa, Hachioji-shi, Tokyo 192-0397, Japan}
\author[0000-0001-9064-160X]{Katsuaki Asano}
\affiliation{Institute for Cosmic Ray Research, The University of Tokyo, 5-1-5 Kashiwanoha, Kashiwa, Chiba 277-8582, Japan}

\begin{abstract}
A giant radio halo (RH) is a diffuse synchrotron emission observed on the scale of megaparsecs (Mpc), typically found in the central region of merging galaxy clusters. 
Its large size and steep spectrum suggest that it originates from the re-energization of an aged population of cosmic ray electrons (CREs), while the secondary leptons produced in the $pp$ hadronic collision of cosmic ray protons (CRPs) may contribute to the emission.
In this study, we investigate the reacceleration model including both primary and secondary CREs, assuming that the primary CRs originate from internal galaxies.
In our new method, we follow the cosmological evolution of each cluster and calculate the energy spectra and one-dimensional spatial distributions of CRs.
The primary CRE model with $\sim 3$ Gyr duration of reacceleration successfully reproduces the statistical properties of the RHs observed in the recent LOFAR survey, as well as the spectrum and profile of the Coma cluster.
The gamma-ray and neutrino emissions produced by reaccelerated CRPs are consistent with the upper limits.
However, if the CRP injection rate is high and the secondary CREs become significant, the model with the required $\sim 3$ Gyr reacceleration overproduces the number of RHs.
The limit on the CRP injection rate, $L_{\rm p} \lesssim 10^{41}$ erg/s, is significantly lower than that expected from the early starburst activity or jets from active galactic nuclei.
This discrepancy requires a revision of either the model of CR supply from galaxies or the turbulent reacceleration model.
\end{abstract}


\keywords{Galaxy clusters (584)}

\section{Introduction} \label{sec:intro}
Galaxy clusters (GCs) are ideal targets for studying the roles of non-thermal processes in the hierarchical formation process of the large-scale structure.
The acceleration of relativistic particles, i.e., cosmic rays (CRs), is associated with various phenomena, including outflows from the active galactic nuclei (AGNs), supernovae in member galaxies, and cosmological shocks and turbulence \citep[e.g.,][]{Norman_1995,Volk_1996,Berezinsky_1997,Bykov_2008,Brunetti_Jones_review}.
Within its gigantic volume, a GC can confine CRs for several billion years (Gyrs) in its magnetized intra-cluster medium (ICM) \citep[e.g.,][]{Volk_1996}.
In the low-density ICM, the energy loss of CRPs is almost negligible, while CREs suffer significant losses due to radiation and Coulomb collisions \citep[e.g.,][]{Sarazin_1999}.
Therefore, the past and recent history of particle accelerations in the cluster environment should be reflected in the distributions of CRPs and CREs, respectively.
\par



The hadronic interactions between CR protons (CRPs) and thermal protons in the ICM produce secondary leptons, gamma rays, and neutrinos.
Observing gamma rays or neutrinos is a promising method to assess the CRP density in the ICM.
However, such emissions have not been detected in most clusters, although some hints of gamma-ray emission have been observed in the Coma cluster \citep[][]{Xi_2018,Abdollahi_2020,Ballet_2020,Adam_2021}.
Recently, \citet{IceCube_2022_stack} reported an upper limit on the neutrino background flux from the Planck Sunyaev-Zel'dovich (SZ) clusters.
\par

In contrast, the observation of radio synchrotron emission from CREs has been successful.
Radio halos (RHs) are diffuse cluster-scale emissions characterized by their large extent and steep spectra, typically found in dynamically disturbed clusters \citep[e.g,][]{vanWeeren_review}.
Recent radio surveys of GCs have revealed various statistical properties of RHs, including their occurrence, correlation with the cluster mass, and the fraction of ultra-steep spectrum RHs \citep[e.g.,][]{Giovannini_1999,Kale_2015,Botteon_2022_LoTSS}.
\par


There are two physical mechanisms for RHs frequently discussed in the literature: reacceleration of relativistic electrons through the interaction with the turbulence in the ICM \citep[e.g.,][]{Brunetti_2001,Peterson_2001,Fujita_2003} and secondary lepton generation through the hadronic $pp$ collisions \citep[e.g.,][]{Dennison_1980ApJ...239L..93D,Blasi_Colafrancesco_1999}. 
Recent discoveries of RHs with very steep radio spectra and diffuse emissions in the peripheral regions of clusters support the reacceleration model \citep[e.g.,][]{Brunetti_2009,BV20,megahalo}.
The reacceleration model is consistent with the statistical properties of RHs \citep[e.g.,][]{Cassano_2006,paperII,Cassano_2023}.
Conversely, the possible detection of GeV gamma rays from the Coma cluster suggests that secondary electrons  might contribute to the RH emission \citep[e.g.,][]{Adam_2021,Keshet_2023,Kushnir_2024}.
Secondary electrons can work as the seed population for the reacceleration model and such a ``hybrid model" can also explain with the RH occurrence \citep[e.g.,][]{Cassano_2012,paperII}.
However, it remains controversial whether CRPs and secondary electrons are essential for RHs.
A detailed examination with radio observations is important for limiting the CRP content in GCs.
\par

We aim to make a comprehensive picture of the distributions of CREs and CRPs in the ICM, and to examine the constraints on the physical parameters imposed by recent multi-messenger observations.
To this end, we revisit the theoretical modeling of RHs, considering the $pp$ collision and the turbulent reacceleration of CRPs and CREs.
We build a 1D spherically symmetric model of the ICM, considering the profiles of thermal gas, magnetic field, turbulence, and CRs.
The novel treatment in our method is the temporal evolution of the radial profiles according to the mass evolution of the DM halo, which is simulated with the so-called merger tree method.
The temporal evolution of the phase space distribution of CRs is calculated with the Fokker--Planck (FP) equation.
The onset of the reacceleration is conditioned by the cluster mergers.
To compare with the RH survey, we calculate the CR evolution and emission for various halos with different merger histories.
\par

In this study, we focus on the injection rate of CRPs in the conventional turbulent reacceleration model \citep[e.g.,][]{Cassano_2013,Brunetti_2017}.
As in previous studies, the radio emission is assumed to be dominated by re-accelerated primary electrons, rather than by secondary electrons.
We assume that the primary CRs are provided from the internal sources, such as galaxies and AGNs, with the redshift evolution of the injection rate proportional to the cluster mass and the star-formation rate density \citep[e.g.,][]{Fang_Murase_2018}. 
We confirm that the reacceleration model can reproduce both the individual and statistical properties of RHs.
However, the CRP injection rate is constrained to be as small as $L_{p} \lesssim 10^{41}$ erg/s. Otherwise, the secondary electrons dominate the radio emission and the reacceleration model overproduces the number of observable RHs.
The CRP injection rate is considerably lower than that expected from the star formation in member galaxies.
This is a question newly raised by this study.
We discuss the possible causes of this low injection rate in the reacceleration model.
\par

In Sect.~\ref{sec:method}, we explain the detail of our 1D time-dependent model. The assumptions about the CR injection and reacceleration can be found in Sect.~\ref{sec:injection} and \ref{sec:Dpp_model}, respectively.
We mainly compare two models: models A and B. The injection of CRPs is completely neglected in the former case.
We consider CRPs and the secondary electrons in Model B, although the primary electrons dominate the emission.
In Sect.~\ref{sec:Coma}, we show that those models can explain the spectrum and the brightness profile of the RH of the Coma cluster.
As shown in Sect.~\ref{sec:stat_RH}, the same models can also explain the statistical properties of the RHs in the recent survey with LOFAR \citep[][]{Botteon_2022_LoTSS}.
The examination of the parameter dependence can be found in Sect.~\ref{sec:discussion}.
Finally, in Sect.~\ref{sec:conclusion}, we summarize our main results.
\par
In the following, we assume a flat $\Lambda$CDM cosmology with the parameters from \citet{Planck18}, where $H_0 = 67$ km s$^{-1}$ Mpc$^{-1}$ ($h = 0.67$ or $h_{70} = 0.96$).

\section{1D model for the evolution of thermal and non-thermal components} \label{sec:method}

We aim to simulate the temporal evolution of the CR distribution in GCs and calculate the emission from various GCs with different mass-evolution histories. 
To follow the cosmological evolution of the dark matter (DM) halo mass, we run a merger-tree simulation using the method of \citet{paperII}.
We adopt a spherically symmetric one-dimensional model for the ICM, utilizing empirical models for the radial profiles of DM and thermal gas. 
We also model the radial profiles of the magnetic field and turbulent strength.
\par

The most significant update to our previous work \citep[][]{paperII} is solving the FP equations for CRPs and CREs (Sect.~\ref{sec:FP}) during the growth of GCs through accretion and mergers.
With this method, we can calculate the emission from various GCs at different redshifts and study both individual and statistical properties of RHs in a consistent manner.
To simplify the discussion in the following sections, we use the fiducial parameters for turbulent reacceleration and focus on constraints on the CR injection rate.
\par

In Sect.~\ref{sec:merger_tree}, we briefly overview the merger tree method.
The profiles for DM and the thermal gas are explained in Sect.~\ref{sec:DM_gas}.
The details of the magnetic field and turbulence models are provided in Sect.~\ref{sec:turb_prof}.
Sect.~\ref{sec:FP} is dedicated to the FP equations for CRs in a mass coordinate, which forms the core of our CR evolution model. 
The assumptions about CR injection and reacceleration are elaborated in Sect.~\ref{sec:injection} and \ref{sec:Dpp_model}, respectively.
The model parameters are summarized in Sect.~\ref{sec:params}.
\par

\subsection{Merger tree of DM halos} \label{sec:merger_tree}
A merger tree describes the history of mergers and mass evolution of the DM halos that eventually merge into a single halo by $z = 0$. In this study, we calculate 5000 trees in total.
We use a Monte Carlo (MC) algorithm to simulate the stochastic growth of the DM halo mass, adopting the halo merger rate and the median mass accretion rate given by \citet{Fakhouri_2010}.
As an initial condition, we generate 5,000 halos at $z = 0$, which are evenly distributed in the logarithmic mass range of $[10^{11}~M_\odot,10^{16}~M_\odot]$.
We follow their mass evolution backward in time (increasing $z$) up to $z =3$, considering the mass decrement due to the merger and the accretion.
\par

We employ the two-body merger approximation, where the descendant halo with the mass of $M_0$ split into two progenitor halos, i.e., $M_0 = M_1 +M_2$.
We define $\xi=M_2/M_1 \leq 1$ as the progenitor mass ratio.
The formula of \citet{Fakhouri_2010} gives the merger rate per halo, $\frac{dN_{\rm m}}{d\xi dz}$, as a function of $M_0$, $\xi$, and $z$.
To distinguish the mergers from the continuous mass accretion, we set the minimum mass ratio for a merger event as $\xi_{\rm min}=10^{-3}$, which is larger than the minimum value of $\xi$ studied in \citet{Fakhouri_2010}. 
\par

The comoving number density of each halo at $z = 0$ is normalized with the Tinker mass function \citep[][]{Tinker_2008}.
The mass of the main halo of $i$-th tree at $z = 0$ is expressed as $M_{z0,i}$.
The comoving number density of a halo in the $i$-th tree, $w_i$, is written as
\begin{equation}\label{eq:tree_weight}
    w_i = \frac{dn}{d\log_{10}M_{z0,i}}\Delta(\log_{10}M_{z0,i}),
\end{equation}
where $dn/d(\log_{10}M_{z0,i})$ is the Tinker mass function at $z = 0$, and $\Delta(\log_{10}M_{z0,i}) = 10^{-3}$ is the bin width for $M_{z0,i}$.
The number of halos of $i$-th tree within a redshift range of $z - (z+\Delta z)$ is
\begin{equation}\label{eq:MT_prob}
    P_i(M_{z0,i},z)\Delta z = w_i\frac{dV_{\rm c}}{dz}\Delta z,
\end{equation}
where $V_{\rm c}$ represents the comoving volume of the Universe.
By summing up $P_i$ for 5000 trees, we confirm that our method can reproduce the Tinker mass function up to $z = 3$. 
For the massive halos ($M > 10^{13}~M_\odot$) at low redshifts ($z \leq 2$), the difference is less than 5\%.
\par

\begin{figure*}
    \centering
    \plottwo{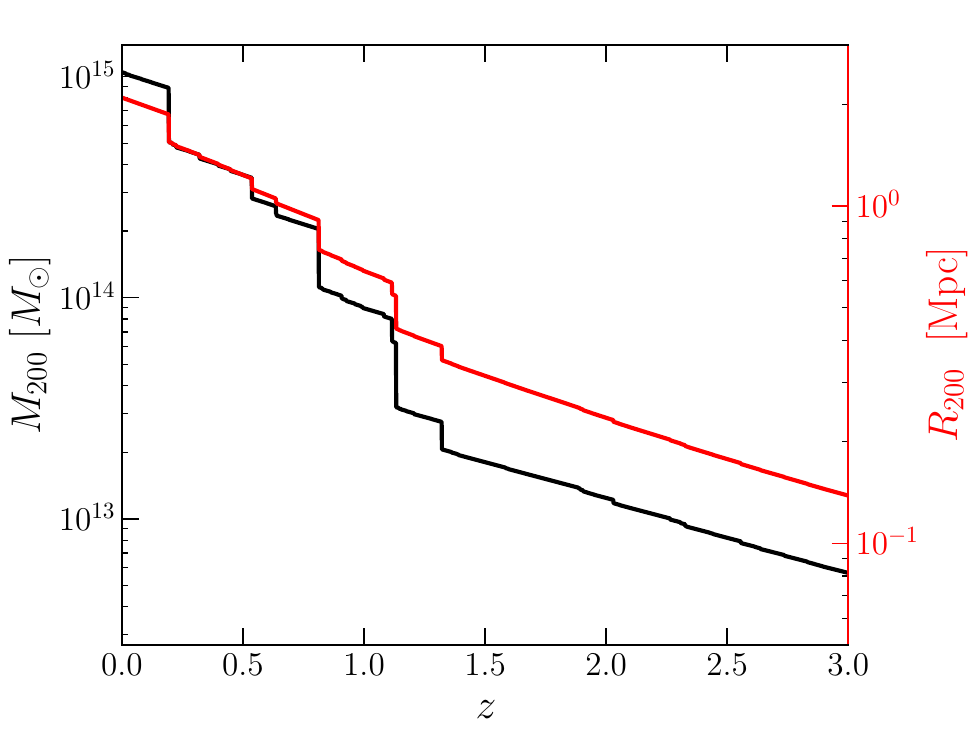}{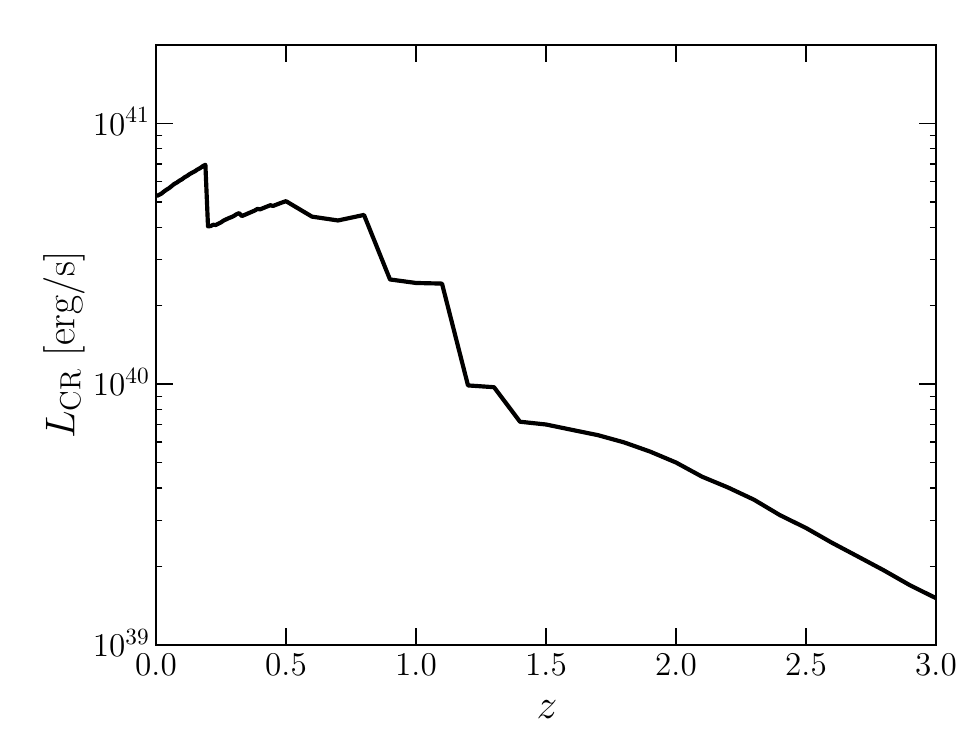}
    \caption{Left: redshift evolution of the mass (black) and the virial radius (red) for an example halo. The merger history of the halo is simulated with the method of \citet{paperII}. The virial radius is measured in the physical coordinate. Right: evolution of CR injection rate for the same halo. The redshift and mass dependence of the injection rate is modeled as Eq.~(\ref{eq:L_CR}). The normalization of the injection rate is constrained from the RH observations in Sect.~\ref{sec:Coma} and Sect.~\ref{sec:stat_RH}.\label{fig:mass_z}}
\end{figure*}

The left panel of Fig.~\ref{fig:mass_z} shows an example of the halo mass evolution simulated with our method.
This halo undergoes multiple merger events for $z<1.5$ and eventually achieves $M_{200} \approx 10^{15}~M_\odot$ at $z = 0$.
The virial radius of the halo, $R_{\Delta}$, where $\Delta=200$ or 500, is the overdensity with respect to the critical mass density of the Universe, is calculated using the DM profile explained in Sect.~\ref{sec:DM_gas}.
The halo experiences a major merger at $z\approx0.2$ with a mass ratio of $\xi = 0.75$ that induces turbulent CR reacceleration in our model.
This simulated halo corresponds to one of the ``RH-hosting" clusters, which exhibits a radio flux observable with the LOFAR sensitivity (see Sect.~\ref{sec:stat_RH}).
\par


\subsection{Modeling DM and gas profiles}\label{sec:DM_gas}

Our method allows us to track the CR distributions over cosmological timescales.
We introduce the Lagrangian mass coordinate and model radial profiles of DM, hot gas, magnetic field, and turbulence, for GCs, where the DM mass evolution is simulated using the merger tree method (Sect.~\ref{sec:merger_tree}). 
Note that our main purpose in this study is the modeling of the statistical properties of GCs, rather than the detailed modeling of individual clusters.
\par


For a given halo mass $M_{\Delta}$ within $R_{\Delta}$ and redshift $z$, the DM density profile is modeled using the Navarro-Frenk-White (NFW) profile \citep[][]{NFW}: 
\begin{equation}\label{eq:NFW}
	\rho_{\rm NFW}(r) =  \frac{\rho_{\rm s}}{(r/r_{\rm s})(1+r/r_{\rm s})^2},
\end{equation}
where $\rho_{\rm s}$ is the central density and $r_{\rm s}$ is the scale radius. It is useful to define the concentration parameter as $c_{\rm vir} \equiv R_{\Delta}/r_{\rm s}$. The central density $\rho_s$ is related to the overdensity parameter as
\begin{equation}
	\rho_{\rm s} =  \frac{\Delta c^3_{\rm vir}}{3[\ln(1+c_{\rm vir})-c_{\rm vir}/(1+c_{\rm vir})]}.
\end{equation}
For the mass and redshift dependence of $c_{\rm vir}$, we adopt the empirical relation reported in \citet{Duffy_2008}:
\begin{equation}
	c_{\rm vir} = A_{\rm con}\left(\frac{M}{M_{\rm pivot}}\right)^{B_{\rm con}}(1+z)^{C_{\rm con}},
\end{equation}
where $A_{\rm con} = 7.85, B_{\rm con} = -0.081,C_{\rm con} = -0.71$, and $M_{\rm pivot} = 2\times10^{12}$ $h^{-1}M_\odot$. 
We use this DM profile to calculate the virial radius $R_{\Delta}$ for various values of $M_{\Delta}$.
In the left panel of Fig.~\ref{fig:mass_z}, we show the redshift evolution of $R_{200}$ associated with the mass evolution shown in the same panel.
\par



Concerning the ICM profile, we adopt the temperature and pressure profiles reported by \citet{McDonald_2014}, based on X-ray observations of 80 GCs selected from the South Pole Telescope survey. 
The deprojected (3D) temperature profile is modeled as \citep[][]{Allen_2001,Vikhlinin_2006,McDonald_2014}:
\begin{equation}\label{eq:T_prof}
	\frac{k_{\rm B}T(r)}{k_{\rm B}T_{500}} = T_0\frac{((x/x_{\rm c})^{a_{\rm cool}}+(T_{\rm min}/T_0))}{(1+(x/x_{\rm c})^{a_{\rm cool}})}\frac{(x/x_{\rm t})^{-a}}{(1+(x/x_{\rm t})^b)^{c/b}},
\end{equation}
where $k_{\rm B}$ is the Boltzmann constant, $x = r/R_{500}$, and $T_{500}$ is the typical temperature for a GC with the mass of $M_{500}$.
This consists of the profiles in two parts: the core region parametrized by a minimum temperature ($T_{\rm min}$), scale radius ($x_{\rm c}$), and shape index ($a_{\rm cool}$), and the outer part expressed by a broken power law with an inner slope ($a$), transition steepness ($b$), outer slope ($c$), and a transition radius ($x_{\rm t}$).
We adopt the mean \citep[including both cool-core and non-cool-core clusters, see][for the detail]{McDonald_2014} profile for low-z sample, where $a_{\rm cool} = 2$, $a = 0$, $b = 2.79$, $c = 0.64$, $T_0=1$, $x_{\rm c} = 0.10$, $x_{\rm t} = 0.40$, and $T_{\rm min}/T_0 = 0.77$. 
As for $T_{500}$, we use the mass--temperature relation reported in \citet{Vikhlinin_2006}:
\begin{equation}\label{eq:T-Mrel}  
	    k_{\rm B}T_{500} = 5~{\rm keV}\left(\frac{E(z)M_{500}}{M_{t}}\right)^{1/\alpha_T},
\end{equation}
where $E(z) = \sqrt{\Omega_{\rm m}(1+z)^3+\Omega_\Lambda}$, $M_{t} = 3.32\times10^{14}$ $h^{-1}M_\odot$, and $\alpha_T = 1.47$. 
Although we consider the increase of turbulent energy after mergers, we ignore the difference in the ICM profile between cool-core and non-cool-core clusters for simplicity. 
\par

The typical pressure of the ICM within $R_{500}$ can be defined as \citep[][]{Nagai_2007} 
\begin{equation}
	P_{500} = n_{\rm g, 500}\times k_{\rm B}T_{500},
\end{equation}
where $n_{\rm g, 500} = 500f_{\rm b}\rho_{\rm cr}/(\mu m_p)$, $f_{\rm b} = \Omega_{\rm b}/\Omega_{\rm m}\approx 0.15$ is the baryon fraction, $\mu\approx0.59$ is the mean molecular weight of the ICM. 
We adopt the so-called generalized NFW profile \citep[][]{Nagai_2007} for the ICM pressure profile with the best fit parameters in \citet{McDonald_2014}:
\begin{equation}\label{eq:gNFW}
	\frac{P(r)}{P_{500}} = f(M_{500})\frac{P_0}{(c_{500}x)^{\gamma_P}(1+(c_{500}x)^{\alpha_{P}})^{\frac{\beta_{P}-\gamma_{P}}{\alpha_{P}}}},
\end{equation}
where $f(M_{500})  = (M_{500}/3\times10^{14}~h_{70}^{-1}M_\odot)^{0.12}$, $P_0 = 4.33$, $c_{500} = 2.59$, $\alpha_{P} = 1.63$, $\beta_{P} = 3.30$, and $\gamma_{P}=0.26$ for low-$z$ clusters. 
The electron number density $n_e(r)$ is calculated consistently with the temperature and pressure profiles as $P(r) = n_e(r)\times k_{\rm B}T(r)$.

\begin{figure*}
    \centering
    \plottwo{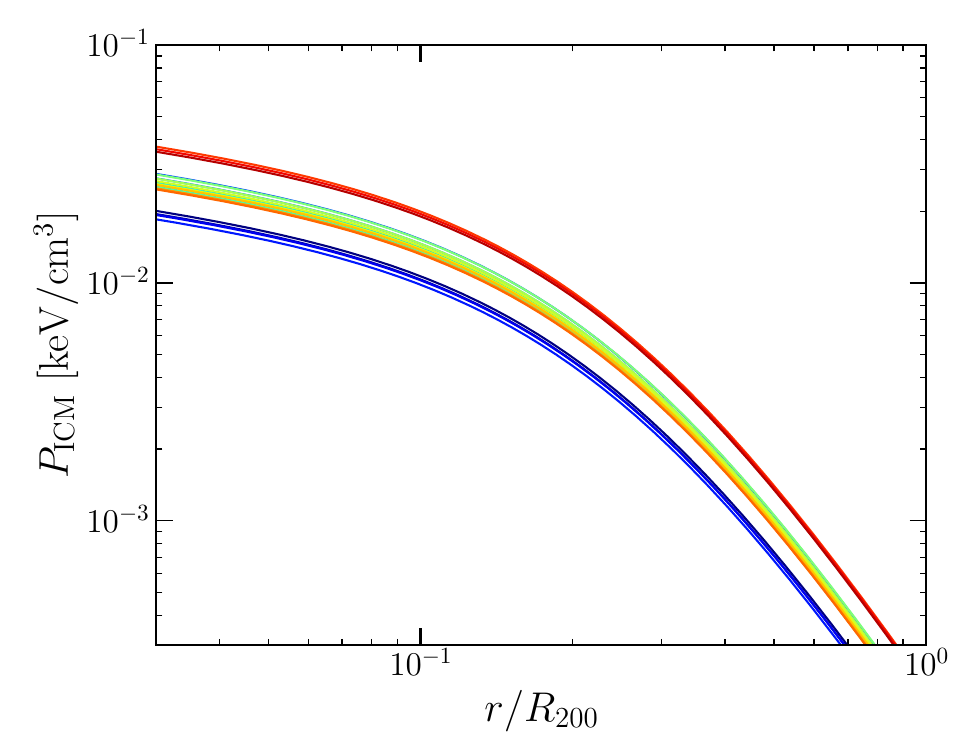}{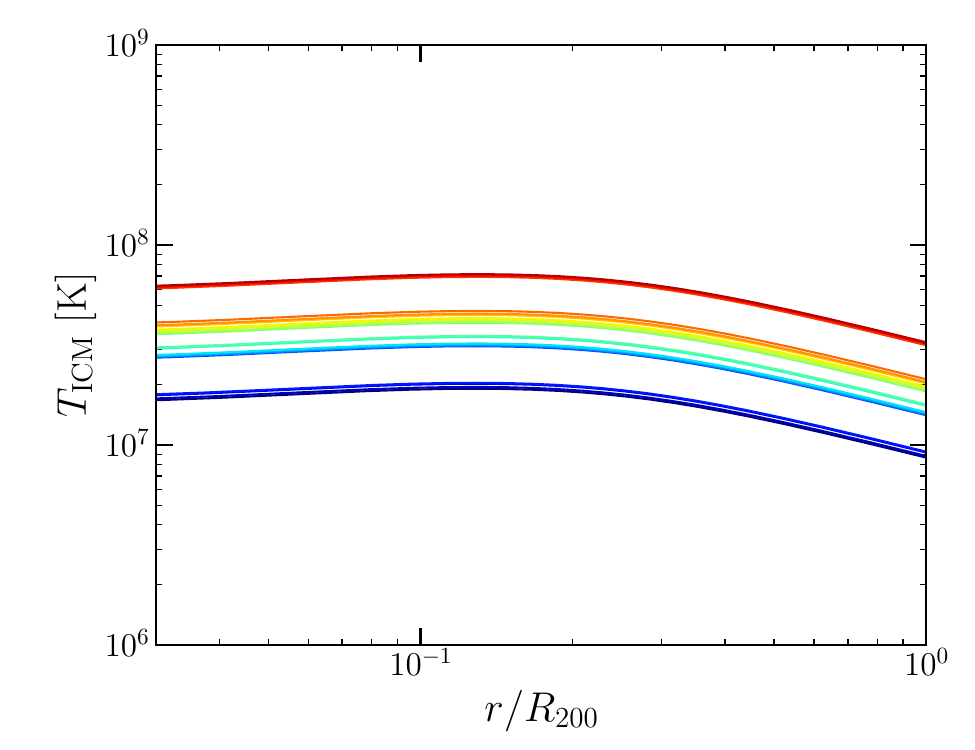}
    \caption{Redshift evolution of the ICM pressure (Left) and the ICM temperature (Right) in $0<z<1$ (every $\Delta z = 0.05$, from blue to red) for the same halo as Fig.~\ref{fig:mass_z}. The radius is normalized with $R_{200}$ at each redshift.
    We adopt the self-similar profiles of Eqs.~(\ref{eq:T_prof}) and (\ref{eq:gNFW}), which are derived from the X-ray observations \citep[e.g.,][]{McDonald_2014}.}
    \label{fig:ICM_prof}
\end{figure*}

Major mergers break the spherically symmetric structure and temporarily disturb the idealized radial profiles modeled above. 
However, for simplicity, we assume that the radial profiles of physical quantities are always modeled with the spherical functions.
Fig.~\ref{fig:ICM_prof} shows an example of the temporal evolution of the ICM thermal pressure and temperature for $0<z<1$ for the same halo in Fig.~\ref{fig:mass_z}.
The normalization of the profiles ($P_{500}$ and $T_{500}$) increase with the mass according to Eqs.~(\ref{eq:T-Mrel}) and (\ref{eq:gNFW}).
Two large gaps between the lines are caused by the abrupt mass increase due to the major mergers at $z = 0.81$ and $z = 0.19$.
Note that the thermal profiles derived from observations of nearby GCs may not be applicable for high redshift ($z>1$) GCs or groups.
However, in our model, the injection and reacceleration of CRs are more efficient for more massive GCs and the results shown in the following sections are not significantly affected by minor changes in the thermal profiles at high $z$.

\subsection{Profiles of magnetic field and turbulence}\label{sec:turb_prof}

Compared to the thermal gas, the profiles of the magnetic field and turbulence in the ICM are less constrained by observations. 
For the magnetic field profile, we adopt the model often used in the RM observations and numerical simulations:
\begin{equation} \label{eq:B_profile}
	B(r) = B_0\left(\frac{n_{e}(r)}{n_0}\right)^{\eta_B}.
\end{equation}
This profile is characterized by two parameters $B_0$ and $\eta_B$. 
In our fiducial models, we adopt the best-fit parameters for the Coma cluster \citep[][]{Bonafede_2010}: $B_0 = 5\mu$G and $\eta_B = 0.5$. 
We also fix $n_{0}= 3\times 10^{-3}$ cm$^{-3}$, which is similar to the central density of Coma cluster \citep[e.g.,][]{Briel_1992}.
Note that the parameter $n_0$ does not mean $n_e(0)$, which evolves with time according to Eqs.~(\ref{eq:T_prof}) and (\ref{eq:gNFW}).
We neglect the redshift dependence of $B_0$.
Recent numerical simulations have shown that the magnetic field in a cluster can be amplified by the dynamo and achieves the saturation with $B\sim1\mu$G by $z\sim2$ \citep[e.g.,][]{Vazza_2018,Steinwandel_2023}.
Recent detection of RHs in distant GCs \citep[][]{DiGennaro_2021NatAs...5..268D,Sikhosana_2024} also suggests rapid magnetic field growth at $z \gtrsim 1$.
\par

To consider the reacceleration of CRs due to the interaction with the turbulence, we introduce the radial profile of the turbulent energy.
The ratio between the turbulent energy and the thermal energy is modeled with the following beta-model-like function:
\begin{equation}\label{eq:turb_prof}
	\frac{\epsilon_{\rm turb}}{\epsilon_{\rm th}}(<r) = A_{\rm t}\left[1+\left(\frac{r/r_{\rm vir}}{B_{\rm t}}\right)^2\right]^{C_{\rm t}},
\end{equation}
where $\epsilon_{\rm turb}$ and $\epsilon_{\rm th}$ are the energy densities of the turbulence and the thermal ICM, respectively, and $A_{\rm t}$, $B_{\rm t}$, and $C_{\rm t}$ are the fitting parameters. 
We assume that the turbulent spectrum $W(k)$ follows the Iroshnikov-Kraichnan scaling, $W(k) \propto k^{-3/2}$, where $k =1/l$ is the wavenumber \citep[][]{Kraichnan_1967}.
To compare with the simulation by \citet{Vazza_2011}, $\epsilon_{\rm turb} \sim kW(k)$ in Eq.~(\ref{eq:turb_prof}) is measured at the fixed scale of $l = 300$ kpc.
\par

As in our previous study \citep[][]{paperII}, we assume that the outer scale of the turbulence $l_{\rm max}$ is proportional to the virial radius of the cluster as $l_{\rm max} \approx 0.5R_{500}$. 
The eddy turnover time at $l_{\rm max}$ can be written as:
\begin{align}\label{eq:teddy}
	t_{\rm eddy}  &=  \frac{l_{\rm max}}{v_{\rm turb}}, \nonumber \\
	&\approx 820~{\rm Myr}\left(\frac{l_{\rm max}}{500~{\rm kpc}}\right)\left(\frac{v_{\rm turb}}{600~{\rm km/s}}\right)^{-1},
\end{align}
where $v_{\rm turb} = \sqrt{2\epsilon_{\rm turb}/\rho_{\rm ICM}}$ is the total turbulent velocity at the injection scale $l_{\rm max}$ and $\rho_{\rm ICM}$ is the mass density of the ICM.
Depending on $R_{500}$, the outer scale $l_{\rm max}$ can be larger or smaller than $l = 300$ kpc, where the turbulent energy is normalized by Eq.~\ref{eq:turb_prof}.
\par

\begin{figure*}
    \plottwo{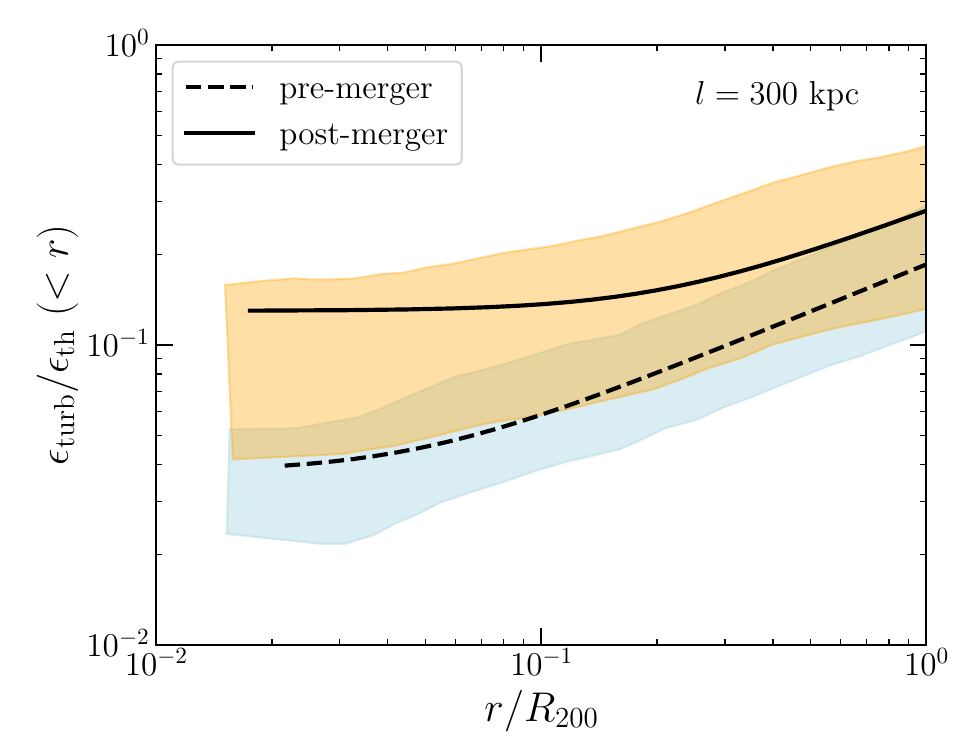}{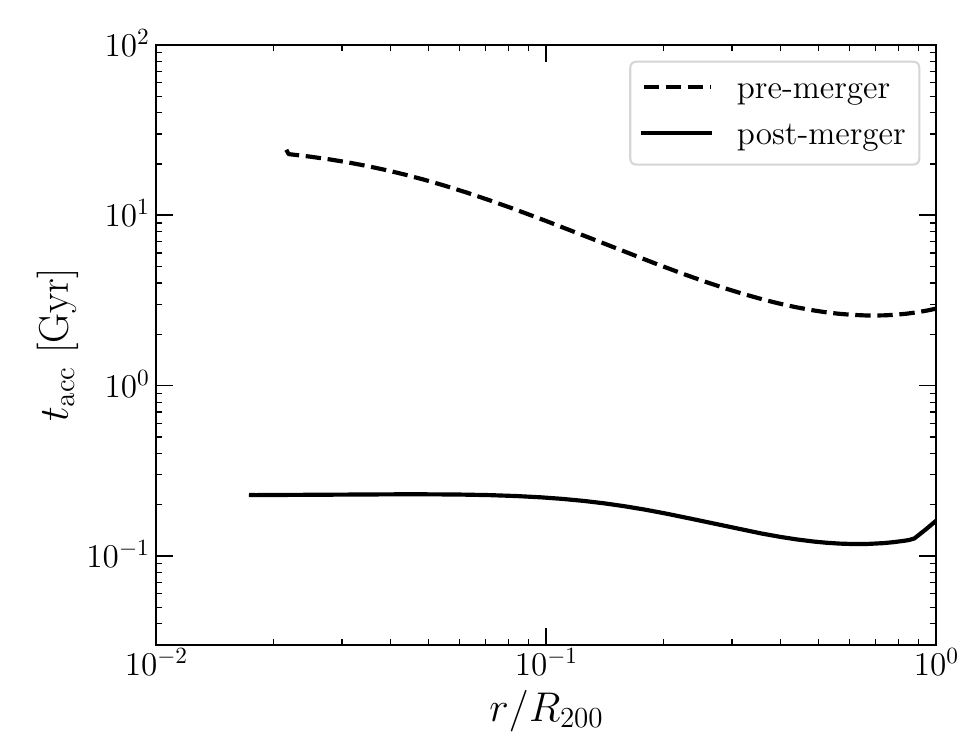}
    \caption{Left: The ratio between the turbulent and thermal energy densities within the radius of $r$ (Eq.~\ref{eq:turb_prof}).  The horizontal axis is normalized by the virial radius $R_{200}$. The solid and dashed lines are post-merger ($\xi = 0.6$) and pre-merger states, respectively. The post-merger turbulence is scaled with $\xi$ (see text for detail).
    The shaded regions show the profiles for the post-merger (orange) and pre-merger (light blue) states found in the simulation by \citet{Vazza_2011}. The turbulent energy is measured at the eddy scale of $l = 300$ kpc.
    Right: The radial profile of the reacceleration time scale $t_{\rm acc}$ calculated with Eq.~(\ref{eq:tacc_TTD_th}). Shown are the results for a cluster with $z = 0.05$ and $M_{500}\approx8\times10^{14}M_\odot$. The post-merger profile (solid line) is compatible with the brightness profile of the Coma RH (see also Sect.~\ref{sec:Coma}).
    The dashed line show $t_{\rm acc}$ for the pre-merger state calculated with $f_{\rm comp}^{\rm pre} = 0.3$. \label{fig:turb}}
\end{figure*}


The turbulence may be always induced by the steady accretion or galaxy motions even in the pre-merger stage. In our model, the turbulence distribution is different for ``pre-merger" and ``post-merger" phases. The two states are expressed with different sets of $(A_{\rm t},B_{\rm t},C_{\rm t})$.
In addition, we take into account the mass ratio dependence of the amplitude $A_{\rm t}$ for the post-merger state.
For the reference value of $\xi = 0.6$, we adopt $(A_{\rm t},B_{\rm t},C_{\rm t}) = (0.14,0.16,0.24)$, which results in $D_{pp}$ similar to the one adopted in \citet{paperII} under the assumption of the TTD reacceleration.
We assume that the total turbulent energy is proportional to the gravitational energy of the merger, $E_{\rm mer} = GM_1M_2/R_{\rm mer}$, where the masses of two progenitor clusters are $M_1 = M_0/(1+\xi)$ and $M_2 = \xi M_0/(1+\xi)$ (Sect.~\ref{sec:merger_tree}), and we choose the radius for evaluation $R_{\rm mer} = R_{\Delta,1} \propto (1+\xi)^{-1/3}$, i.e., the virial radius of the major progenitor halo \citep[e.g.,][]{Cassano_Brunetti_2005}.
In this case, the amplitude of the turbulence scales as $A_{\rm t} \propto \xi/(1+\xi)^{5/3}$.


In Fig.~\ref{fig:turb}, we show the post-merger profile for $\xi = 0.6$ with the solid line.
We can compare this model line with the orange region that shows the post-merger profile in the simulation by \citet{Vazza_2011}. 
\par

\begin{table}
	\centering
	\caption{parameters for the turbulence profile Eq.~(\ref{eq:turb_prof}) \label{tab:turb_prof}}
	\begin{tabular}{cccc}
	\hline
	&$A_{\rm t}$&$B_{\rm t}$&$C_{\rm t}$ \\
	\hline
	pre-merger & 0.036 & 0.037 &  0.26  \\
	post-merger ($\xi = 0.6$) & 0.14 &  0.16 & 0.24 \\
	\hline
	\end{tabular}
\end{table}

On the other hand, the turbulent profile in the pre-merger state can hardly be constrained from the modeling of RHs.
In this work, we adopt the mean profile found in the simulation in \citet{Vazza_2011}, which can be fitted with $(A_{\rm t},B_{\rm t},C_{\rm t}) = (0.036,0.037,0.26)$.
The pre-merger profile is shown with the dashed line in Fig.~\ref{fig:turb}.
The values of $(A_{\rm t},B_{\rm t},C_{\rm t})$ are summarized in Tab.~\ref{tab:turb_prof}.
\par

In our calculations, we assume that the turbulent profile of a cluster is initially in the pre-merger state.
When the cluster experiences a merger with a mass ratio of $\xi > \xi_{\rm th}$, where $\xi_{\rm th}$ is the threshold value, the cluster transitions to the post-merger state.
The cluster then recovers the relaxed state after a duration of $T_{\rm dur}$, where $T_{\rm dur}$ is treated as a parameter. 
The parameter $T_{\rm dur}$ represents the lifetime of the merger-induced turbulence, which also corresponds to the duration of the merger-induced reacceleration.
In our calculation, $T_{\rm dur}$ is normalized by $t_{\rm eddy}$ (Eq.~(\ref{eq:teddy})).
For simplicity, we do not consider the gradual decay of turbulent energy in the post-merger state. 
\par

As will be explained in Sect.~\ref{sec:Dpp_model}, we consider the transit-time damping (TTD) as the mechanism of stochastic acceleration of CRs in the turbulent ICM, which is conventionally adopted in the literature \citep[e.g.,][]{Cassano_2006,BL11,Cassano_2023}.
The TTD resonance is an interaction between charged particle and the compressible component of turbulence, while the profile in Eq.~(\ref{eq:turb_prof}) corresponds to the total energy of the turbulence. 
We specify the fraction of the energy density of compressible mode to the total turbulent energy by introducing a parameter $f_{\rm comp}$.
The compressible fractions for the pre-merger and post-merger states are expressed as $f_{\rm comp}^{\rm pre}$ and $f_{\rm comp}^{\rm post}$, respectively.
\par

Note the degeneracy in the parameters, $(A_{\rm t},B_{\rm t},C_{\rm t})$, and $f_{\rm comp}$.
In this work, we assume $f_{\rm comp}^{\rm post} \approx 1$, i.e., the post-merger turbulent energy is dominated by the compressible component.
As for the pre-merger state, we set $f_{\rm comp}^{\rm pre} = 0.3$ to reduce the effect of the reacceleration.
Re-acceleration in the pre-merger state can be important for the CRP spectrum, so we test the case of $f_{\rm comp}^{\rm pre} = 0$ in Sect.~\ref{sec:modelC}.
\par



\subsection{FP equations for relativistic particles in a mass coordinate}\label{sec:FP}
We solve FP-type equations to follow the temporal evolution of CR distribution taking into account injection, cooling, diffusion, and reacceleration. 
In our previous work \citep[][]{paperII}, we assumed a simplified relation between the cluster mass and the radio luminosity.
However, in the current work, we solve the FP equation for each cluster to estimate the non-thermal emissions.
We assume that the phase space distribution of CRs, $f(\bm{x},\bm{p},t)$, is isotropic in the momentum space and geometrically spherical, i.e., it can be written as $f(r,p,t)$. 
The particle distribution per unit momentum per unit radial distance is defined as
\begin{equation}\label{eq:N}
	N_s(r,p,t) \equiv (4\pi r^2)(4\pi p^2) f_s,
\end{equation}
where $s = e,p$ stands for the particle species. One can find FP equations for $N_s$ in the literature \citep[e.g.,][]{paperI,Brunetti_2017}.
\par

In this study, we consider the growth of the cluster halo and the cosmological expansion of the Universe. In this case, it is useful to introduce the mass coordinate:
\begin{equation}\label{eq:mass_co}
	m(r) = \int_0^{r}dr' 4\pi r'^{2} \rho(r'),
\end{equation}
where $\rho$ is the mass density, and $r$ and $r'$ are the cluster radii measured in the physical coordinate. 
The physical and comoving coordinates are related by
\begin{eqnarray}\label{eq:r_com}
    r = a(z)r_{\rm c},
\end{eqnarray}
where $a(z) = 1/(1+z)$ is the scale factor. 
We use $\tau$ for the time in the mass coordinate and consider the transformation of the coordinate from $(t,r_{\rm c})$ to $(\tau, m)$, where $\tau = t$.
We consider the thermal gas as a background matter in Eq.~(\ref{eq:mass_co}), i.e., $\rho = \rho_{\rm ICM}$. 
We prepare 256 mass bins for the gas mass in the range of $0.005M_{z3} < m < 0.5M_{z0}$, where $M_{z3}$ and $M_{z0}$ are the halo virial masses ($M_{200}$) at $z = 3$ and $z = 0$, respectively.
The bins are evenly spaced in the logarithmic space.
At each time step, the profiles of the thermal gas, magnetic field, turbulence, and CRs are calculated for the bins in $m < M_{200}$, i.e., $r < R_{200}$.
As the halo mass increases due to accretion and mergers, the number of bins involved in the calculation increases.
\par

We define the distribution of CRs in each mass bin as \citep[e.g.,][]{Jubelgas_2008}
\begin{equation}\label{eq:N_til}
	\tilde{N}_s(m,p,\tau) \equiv  4\pi p^2 \frac{1}{\rho_{\rm ICM}(r_{\rm c})}f_s(r_{\rm c},p,t),
\end{equation}
With this definition, $\tilde{n}_s = \int dp \tilde{N}_s$ becomes the specific number density of CRs. 
Using the FP equation in \citet{paperI} and Eqs.~(\ref{eq:mass_co}), (\ref{eq:r_com}) and (\ref{eq:N_til}), one can find the FP equation in the mass coordinate in the form of
\begin{align}\label{eq:FP_mass}
	\frac{\partial \tilde{N}_s}{\partial \tau} &= \frac{\partial}{\partial p}[b_s\tilde{N}_s]-\frac{\partial}{\partial p}\left[\frac{\tilde{N}_s}{p^2}\frac{\partial}{\partial p}(pD_{pp})\right]+\frac{\partial^2}{\partial p^2}\left[D_{pp}\tilde{N}_s\right]\nonumber \\
	&\; +\frac{1}{3}\frac{1}{\rho_{\rm ICM}}\frac{\partial\rho_{\rm ICM}}{\partial \tau}\frac{\partial}{\partial p}[p\tilde{N}_s]  \nonumber \\
    &\; +16\pi^2\frac{\partial}{\partial m}\left[a^4r^4_{\rm c}D_{rr}\rho_{\rm ICM}\left(\frac{\partial \rho_{\rm ICM}}{\partial m}\tilde{N}_s + \rho_{\rm ICM}\frac{\partial \tilde{N}_s}{\partial m}\right)\right] \nonumber \\
    & \; -\frac{\tilde{N}_s}{t_{\rm loss}}+\tilde{Q}_s^{\rm inj}(m,p,\tau),
\end{align}
where $b_{\rm s} = \dot{p}_s$ represents the energy loss rate, $D_{pp}$ is the momentum diffusion coefficient due to the turbulent reacceleration, $D_{rr}$ is the radial diffusion coefficient, $-N_s/t_{\rm loss}$ is the sink term, and $\tilde{Q}_s^{\rm inj}$ is the injection term of primary CRs. 
Concerning the equation for $\tilde{N}_p$, we adopt the $pp$ collision timescale for $t_{\rm loss}$. 
The sink term is not included in the equation for $\tilde{N}_e$. 
Concerning the energy loss rate and the spatial diffusion coefficient, we use the same model as \citet{paperI}, where we consider the cooling of CRE due to the Coulomb collision and the radiative (synchrotron and inverse-Compton) cooling, the Coulomb loss of CRPs, and the spatial diffusion due to the pitch-angle scattering \citep[e.g.,][]{Murase_2013,Fang_Olinto_2016}. 
The spatial diffusion coefficient for GeV CRs is approximately $D_{rr}\sim 10^{29.5}$ cm$^2$/sec. Considering uncertainty in the CR transport in the ICM, we test the case of a constant diffusion coefficient, $D_{rr} = 10^{31}$ cm$^2$/s in Sect.~\ref{sec:modelD}.
\par

The fourth term on the right-hand side of Eq.~(\ref{eq:FP_mass}) represents the adiabatic compression or expansion.
The adiabatic term is proportional to $\dot{\rho}_{\rm ICM}/\rho_{\rm ICM}$ measured in each mass bin.
Note that $\rho_{\rm ICM}$ at a certain mass bin evolves with the halo mass according to Eqs.~(\ref{eq:T_prof}) and (\ref{eq:gNFW}).
The adiabatic CR cooling/heating in the central region of the cluster occurs on a timescale of $\sim1/H(z)$, where $H(z) = H_0E(z)$, and is generally less efficient than other processes, such as diffusion and radiative cooling.
\par


The above FP equation is solved for the most massive progenitor halo in each tree. 
In this study, we neglect the CR acceleration due to the diffusive shock acceleration or the adiabatic compression at the accretion shock. 
The CR density in the newly-added mass bin resulting from mass accretion is set to zero.
At the merger event, the CRs in the two progenitor clusters are mixed.
For simplicity, this effect is treated by multiplying a factor of $(1 + \xi)$ to the CR number density, $\rho_{\rm ICM}\tilde{N}_s$, of the main (heavier) progenitor.
In our model, turbulent reacceleration plays a more significant role than the CR supply from the companion regarding the increase of the CR energy density.
\par

For the injection of CREs, we consider both primary CREs and secondary CREs from the $pp$ collision. 
The production rate of the secondary particles is calculated in the same manner as \citet{paperI}.
The photohadronic interactions of CRPs are neglected in this study since they are important only for the ultra-high energy CRs ($E_p > 10^{18}$ eV).
We adopt the fully-implicit Chang-Cooper method to solve the above FP equation \citep[][]{CC70,Park_Petrosian_1995}. 
In the following sections, we omit the subscript, ${\rm c}$, for the comoving coordinate.
\par

\subsection{Injection of primary CRs}\label{sec:injection}

The injection of primary CRs is represented by the injection term $\tilde{Q}_s^{\rm inj}(m,p,\tau)$ in the FP equation (Eq.~\ref{eq:FP_mass}). 
The primary CRs would be accelerated at cosmological shocks, supernovae in the member galaxies, or the AGNs, although it is not clear which is the dominant source \citep[e.g.,][]{Norman_1995,Murase_2008,Fang_Murase_2018}.
In this work, we specifically consider the case of CR injection from internal sources including galaxies and their AGNs.
For simplicity, we assume that the primary CRE and CRP follow single power-law spectra with the index of $\alpha_{\rm inj}$. 
\par

We consider the redshift and mass dependence of the CR injection rate.
For the redshift dependence, $\mathcal{E}(z)$, we assume that the injection rate is proportional to the cosmic star-formation rate density, following the previous studies \citep[e.g.,][]{Murase_2013,Fang_Murase_2018,Hussain_2021}:
\begin{align}\label{eq:CR_evolution}
 \mathcal{E}(z) = C_z\frac{(1+z)^{p_1}}{1+[(1+z)/(1+z_0)]^{p_2}},
\end{align}
where $p_1 = 2.9$, $p_2 = 5.6$, and $z_0  = 1.9$ \citep[][]{Madau_Dickinson_2014}. 
The normalization $C_z$ is determined by the condition $\mathcal{E}(0) = 1$.

For the mass dependence, we test the case of $L_{\rm CR}\propto M^1$, as usually assumed in the literature \citep[e.g.,][]{Murase_2013,Zandanel_2015,Fang_Murase_2018}. 
The injection luminosity is normalized at $z = 0$ and $M_{500} = 10^{15}~M_\odot$ with $L_{s}^{\rm inj}$, which is treated as a parameter.
Summarizing above, the CR injection luminosity in a GC with the mass $M_{500}$ at $z$ can be expressed as
\begin{equation}\label{eq:L_CR}
	L_{s}  = L^{\rm inj}_{s}\mathcal{E}(z)\left(\frac{M_{500}}{10^{15}~M_\odot}\right).
\end{equation}


In order to specify the functional form of $\tilde{Q}_s(m,p,\tau)$ in Eq.~(\ref{eq:FP_mass}), we need the radial profile of the CR injection.
In this work, we consider the injection rate density proportional to the number density of galaxies in the GC. 
We adopt the 3D King profile \citep[][]{King_1962,King_1972,Adami_1998} of the radial distribution of galaxies:
\begin{equation}\label{eq:King}
	n_{\rm gal}(r)  = \frac{n_{\rm gal,0}}{\left[1+\left(\frac{r}{r_{\rm core}}\right)^2\right]^{3/2}},
\end{equation}
where $n_{\rm gal,0}$ is the central number density and $r_{\rm core}$ is the core radius. 
We adopt the empirical relation between $r_{\rm core}$ and $R_{200}$ found in the SDSS survey \citep[][]{Popesso_2007}: $r_{\rm core} \approx 0.22\times R_{200}$. 
\par

The injection term in the FP equation (Eq.~(\ref{eq:FP_mass})) can be written as:
\begin{align}\label{eq:Q_AGN}
	\tilde{Q}_s(m(r),p,\tau(z)) &=  C_s^{\rm inj} p^{-\alpha_{\rm inj}}\left[1+\left(\frac{r}{r_{\rm core}}\right)^2\right]^{-3/2} \nonumber \\
 &\; \times \mathcal{E}(z)\left(\frac{M_{500}}{10^{15}~M_\odot}\right),
\end{align} 
where $C^{\rm inj}_s$ is determined from the normalization parameter $L^{\rm inj}_{s}$.
The injection luminosity $L_s$ (Eq.~(\ref{eq:L_CR})) and $\tilde{Q}_s$ (Eq.~(\ref{eq:Q_AGN})) are related as
\begin{equation}
    L_s =  \int_0^{M_{200}}dm \int_{p_{0,s}} dp E_s(p) \tilde{Q}_s(m,p,\tau),
\end{equation}
where the particle energy expressed with momentum $p$, $E_s(p) = m_sc^2\sqrt{(p/m_sc)^2+1}$, and $p_{0,s}$ is the minimum momentum of the primary CRs. 
We adopt $p_{0,p}/(m_pc)=1$ for CRPs. On the other hand, we adopt $p_{0,e}/(m_ec)=300$, which corresponds to the momentum where the cooling time of CREs takes maximum \citep[see Fig.~1 of][]{paperI}.
In Sect.~\ref{sec:modelE}, we examine the case of $p_{0,e}/(m_ec) = 1$ and find that such a model overproduces the number of RHs with large radio fluxes. 
We neglect the mass or redshift dependence of the parameters, $\alpha_{\rm inj}$ and $p_{0,s}$.
\par



\subsection{Model for turbulent reacceleration}\label{sec:Dpp_model}

In this work, we consider the reacceleration by the TTD resonance between the compressible waves and CRs \citep[e.g.,][]{BL07}, although there are several alternatives of the reacceleration mechanisms assumed for cluster-scale diffuse radio emission \citep[e.g.,][]{Fujita_2015,BL16}. 
In the ``collisionless ICM" model of the TTD acceleration \citep[e.g.,][]{Brunetti_2016}, the acceleration timescale ($t_{\rm acc} = p^2/4D_{pp}$) depends on the turbulent Mach number $M_{\rm s}$, the size of the outer scale $l_{\rm max}$, and the sound speed $c_{\rm s}$:
\begin{align}\label{eq:tacc_TTD_th}
	t_{\rm acc} & \approx  320~{\rm Myr}~\left(\frac{f_x}{0.02}\right)^{-1}\left(\frac{M_{\rm s}}{0.5}\right)^{-4} \nonumber \\
    & \; \left(\frac{l_{\rm max}}{0.3~{\rm Mpc}}\right)\left(\frac{c_{\rm s}}{1000~{\rm km/s}}\right)^{-1},
\end{align}
where 
\begin{align}\label{eq:fx}
	f_x &= \frac{c_{\rm s}}{c}\int_0^1d\mu\frac{1-\mu^2}{\mu}\Theta\left(1-\frac{c_{\rm s}}{c\mu}\right)\left[1-\left(\frac{c_{\rm s}}{c\mu}\right)^2\right], \nonumber \\
	& = x\left(\frac{x^4}{4}+x^2-(1+2x^2)\ln x-\frac{5}{4}\right)\approx 0.02,
\end{align}
and $x = c_{\rm s}/c$.
One can evaluate the turbulent velocity $v_{\rm turb}$ at each position of the mass coordinate $m(r)$ using the turbulent energy density $\epsilon_{\rm turb}$ as $v_{\rm turb}= \sqrt{2\epsilon_{\rm turb}/\rho_{\rm ICM}}$.
The turbulent Mach number appearing in Eq.~(\ref{eq:tacc_TTD_th}) is defined as $M_{\rm s} = v_{\rm turb}/c_{\rm s}$.
The outer scale is fixed to $l_{\rm max} = 0.5R_{500}$ (see Sect.~\ref{sec:turb_prof}).
\par

The turbulent energy density given by Eq.~(\ref{eq:turb_prof}) may overestimate the acceleration efficiency at large radii, because the gas in those regions formed at a late epoch of the cluster evolution, and the turbulent cascade may have not reached the dissipation (cut-off) scale. 
Following \citet{paperII}, we introduce a cut-off term as a function of $r$: $D_{pp}\propto \exp(-r/R_{500})$.
\par

In the above model, CRs resonate with a wave at the cut-off scale of the turbulence $l_{\rm cut} = 1/k_{\rm cut}\sim1~{\rm kpc}$, which is determined by the TTD resonance of the thermal electrons. 
The acceleration timescale of Eq.~(\ref{eq:tacc_TTD_th}) is comparable to the cooling time of radio-emitting electrons in $\sim\mu$G magnetic field.
The steep spectral nature of RHs can be explained by the balance between the reacceleration and the cooling \citep[e.g.,][]{Cassano_2010_LOFAR}.
Note that $t_{\rm acc}$ does not depend on the particle momentum $p$.
The second-order Fermi acceleration of this type is often termed as the ``hard-sphere" acceleration.
\par

In the right panel of Fig.~\ref{fig:turb}, we show the radial profile of $t_{\rm acc}$ for the pre-merger (dashed) and post-merger (solid) states.
We assume a GC at redshift $z\approx 0.05$ and $M_{500} = 8\times 10^{14}~M_\odot$.
The acceleration timescale gradually decreases with $r$ due to the radially increasing trend of $\epsilon_{\rm turb}/\epsilon_{\rm th}$.
The cut-off term $D_{pp}\propto \exp(-r/R_{500})$ increases the acceleration timescale at the edge of GCs.
The reacceleration timescale in the pre-merger state is several Gyrs due to the small $f_{\rm comp}^{\rm pre}$.
Such a weak reacceleration is not effective for CREs due to the strong radiative cooling. 
On the other hand, CRPs can be re-accelerated even in the pre-merger state (see Sect.~\ref{sec:modelC}).
\par

As explained in Sect.~\ref{sec:turb_prof}, the turbulent energy density $\epsilon_{\rm turb}$ and the largest scale of the turbulence $l_{\rm max}$ are proportional to the thermal energy density and the virial radius of the GC, respectively.
As one can approximate $f_x \propto c_{\rm s}$, the parameter dependence of the TTD acceleration timescale becomes $t_{\rm acc} \propto M_{\rm s}^{-4}R_{500}c_{\rm s}^{-2}$ (Eq.~(\ref{eq:tacc_TTD_th})). 
Since we fix the fraction between the thermal and turbulent energy densities with Eq.~(\ref{eq:turb_prof}), $M_{\rm s}$ is independent of the mass or redshift. 
Using the definition of the virial radius, one finds $l_{\rm max}(M_{500},z)\propto M_{500}^{1/3}E(z)^{-2/3}$.
As explained in Sect.~\ref{sec:DM_gas}, we adopt the temperature--mass relation found in the X-ray observation. 
The typical sound speed of the ICM scales as $c_{\rm s}(M_{500},z) \propto M_{500}^{1/3}E(z)^{1/3}$. Thus, the mass and redshift dependence of $t_{\rm acc}$ can be expressed as
\begin{equation}\label{eq:tacc_scale}
	t_{\rm acc}(M_{500},z) \propto M_{500}^{-1/3}E(z)^{-4/3}.
\end{equation}
For a fixed mass, the reacceleration becomes more efficient at high redshift because the physical scale of the cut-off scale becomes smaller and the sound velocity becomes faster. Similarly, the eddy turnover time can be expressed as 
\begin{equation}
	t_{\rm eddy}(z) \propto E(z)^{-1},
\end{equation}
and is independent of the mass. 
As the fraction between the duration and the timescale of the reacceleration, $T_{\rm dur}/t_{\rm acc} \propto t_{\rm eddy}/t_{\rm acc}$, increases, the spectral evolution due to the reacceleration becomes more and more apparent \citep[][]{paperI}. 
This fraction $t_{\rm eddy}/t_{\rm acc}$ increases with the mass, while it is nearly independent of $z$ in $z < 1$. 
Because the halo mass generally decreases with redshift,  $t_{\rm eddy}/t_{\rm acc}$ decreases with redshift, which means that the merger-induced reacceleration is more efficient in low $z$. 
\par


The expression of Eq.~(\ref{eq:tacc_TTD_th}) is derived under the assumption of the long wavelength limit of the plasma wave \citep[e.g.,][]{Barnes_Scargle_1973}. 
The model of TTD acceleration becomes invalid when the Larmor radius of CRs becomes comparable to the cut-off scale $l_{\rm cut}$. 
To incorporate this effect, we introduce a cut-off in the expression of $D_{pp}$: $D_{pp}\propto \exp(E(p)/E_{\rm max})$, where $E_{\rm max} = 2\pi eB/k_{\rm cut}$.
\par

Summarizing above, the functional form of the momentum diffusion coefficient becomes
\begin{align}\label{eq:Dpp_model}
	D_{pp}(m,p,z) &= \frac{p^2}{4t_{\rm acc}(m(r),M_{500},z)} \nonumber \\
 & \times \exp\left(-\frac{r_{\rm c}}{R_{500}(z)}\right)\exp\left(-\frac{E(p)}{E_{\rm max}(m(r),z)}\right),
\end{align}
where $t_{\rm acc}(m(r_{\rm c}),z)$ is calculated with Eq.~(\ref{eq:tacc_TTD_th}) and the turbulent profile of Eq.~(\ref{eq:turb_prof}). The radial profile of the acceleration timescale is shown in Fig.~\ref{fig:turb}.  
When the efficiency of the reacceleration is very large, the turbulence should be damped by the back reaction of the reacceleration. 
We artificially set $D_{pp} = 0$ when the CR energy density exceeds the turbulent energy density. 
Thus, the CR energy density is limited by the turbulent energy density during the reacceleration. 
However, under a reasonable assumption on the parameters, $\epsilon_{\rm CRP} < \epsilon_{\rm turb}$ is satisfied in most of GCs.
\par

\subsection{Model parameters}\label{sec:params}
\begin{table*} 
	\centering
	\caption{fiducial values for the parameters in our reacceleration model \label{tab:params}}
	\begin{tabular}{cccc}
	\hline
	& parameter & symbol & fixed value\\
	\hline
	Magnetic field & Central field strength & $B_0$ & 5~$\mu$G \\
	& Index of the radial profile (Eq.~(\ref{eq:B_profile}))& $\eta_{B}$ & 0.5 \\
	\hline
	Turbulence & Outer (injection) scale of the turbulence & $L$ & $0.5R_{500}$ \\
	& Radial profile of  turbulence (Eq.~(\ref{eq:turb_prof})) &  Tab.~\ref{tab:turb_prof}\\
        & Threshold mass ratio for the reacceleration & $\xi_{\rm th}$ & 0.2 \\
	& Duration of the post-merger state & $T_{\rm dur}$ & 4$t_{\rm eddy}$ \\
	& Energy fraction of the compressible mode:\\
	& in post-merger state & $f_{\rm comp}^{\rm post}$ & 1 \\
	& in pre-merger state &$f_{\rm comp}^{\rm pre}$ & 0.3 \\
	
	\hline
	CR injection  
        & Spectral index of primary CRs & $\alpha_{\rm inj}$ &  2.2\\
        & Minimum momentum of primary CRE & $p_{\rm e}^{\rm min}$ & 300 \\
	\hline
 	\end{tabular}
\end{table*}
In this section, we summarize the model parameters. 
To organize our calculations so that they are easy to interpret, we need to fix some parameters.  
In this paper, we focus on the injection of primary CRs in the conventional reacceleration model.
Thus, we adopt the reacceleration parameters similar to the ones found in the literature \citep[e.g.,][]{Cassano_2006,Brunetti_2017}.
We also fix the parameters for the DM and thermal gas profiles (Sect.~\ref{sec:DM_gas}). 
The parameters for the magnetic field, turbulence, and the CR injection are summarized in Tab.~\ref{tab:params}.
\par


The radial profile of the turbulence can be expressed with Eq.~(\ref{eq:turb_prof}) and the three parameters are listed in Tab.~\ref{tab:turb_prof}.
The transition between the pre-merger and the post-merger states can be expressed with two parameters: $\xi_{\rm th} $ and $T_{\rm dur}$.
The post-merger turbulent energy depends on the merger mass ratio (Sect.~\ref{sec:turb_prof}). 
The effect of the reacceleration during the pre-merger state is studied with the parameter $f_{\rm comp}^{\rm pre}$.
\par


In \citet{paperII}, the duration of the acceleration is assumed as $T_{\rm dur} = t_{\rm eddy}\approx 0.7$ Gyr. In that case, $\xi_{\rm th}$ is required to be as small as $\sim0.01$ in order to explain the occurrence of RHs. 
However, a merger with such a small mass ratio may not result in the clear dynamical disturbance of the ICM observed in the RH-hosting GCs \citep[e.g.,][]{Cuciti_2023}.
In our fiducial models, we adopt $\xi_{\rm th} = 0.2$ and $T_{\rm dur}= 4t_{\rm eddy}\approx 3$ Gyr, following \citet{Cassano_2016}. The consistency between this model and the RH occurrence will be confirmed in Sect.~\ref{sec:stat_RH}.
In Appendix~\ref{app:T_dur}, we test the case of $T_{\rm dur}= 2t_{\rm eddy}\approx 1.5$ Gyr to study how the duration time affects the estimate of the CR injection rate.
\par


In the internal-source model for the CR injection, the injection luminosity is proportional to the halo mass $M_{500}$ and its redshift evolution is proportional to that of the star-formation rate (Eq.~(\ref{eq:CR_evolution})), irrespectively of the post/pre-merger phases.
We fix $\alpha_{\rm inj} = 2.2$ for the spectral index and treat the normalization factor $L_{s}^{\rm inj}$ as a free parameter.
The minimum momentum of the primary CRE is fixed to $p_{0,e}/(m_ec) = 300$ in our fiducial model.
This choice is suitable in explaining the RH number count in the LOFAR survey (Sect.~\ref{sec:stat_RH}).
The case of $p_{0,e}/(m_ec) = 1$ is discussed in Sect.~\ref{sec:modelE}.
\par


In the following sections, we discuss the constraints on $L^{\rm inj}_{s}$ by modeling the observed RHs. 
We compare two models: Models A and B. Model A corresponds to a pure primary-electron reacceleration model, where we assume $L_{p}^{\rm inj}=0$ erg/s. In Model B, we consider the injection of primary CRPs and the secondary production though the $pp$ collision. The CRP injection luminosity is taken to be sufficiently small and the radio emission in $\nu\lesssim$10GHz is dominated by the primary electrons.
\par



\section{Coma cluster}\label{sec:Coma}
Before the statistical discussion for RHs, we start with the modeling of the Coma cluster and evaluate the typical values of the CR injection rates $L_{\rm e}^{\rm inj}$ and $L_{\rm p}^{\rm inj}$.
This demonstration would ensure our parameter sets in this paper, as the Coma cluster is one of the most studied examples of the nearby massive cluster hosting an RH.
The redshift of the Coma cluster is $z_{\rm obs} = 0.0231$ \citep[e.g.,][]{Struble_Rood_1999}.
The Coma RH may not be a ``typical" example of RHs, because the radio brightness of the Coma RH is almost one order of magnitude fainter than other RHs with similar physical sizes as reported in \citet{Murgia_2024}.
In this work, we assume that the cluster is at the early stage of the turbulent reacceleration and the emission growth is now in progress.
Similar assumptions have been made in previous studies \citep[e.g.,][]{BL07,Brunetti_2017,Pinzke_2017,paperI}.
\par


During the reacceleration phase, the CR spectra and the resulting emission evolves significantly in the timescale of a few 100 Myrs.
The estimate on $L_{\rm e}^{\rm inj}$ and $L_{\rm p}^{\rm inj}$ depends on the redshift of the merger which initiates the reacceleration, $z_{\rm st}$.
Following the previous studies, we assume $z_{\rm st} \approx 0.06$, which is approximately $\Delta T_{\rm dur} = 500$ Myr before the observed redshift \citep[e.g.,][]{Brunetti_2017}.
Because of the limited range of $z_{\rm st}$ as $0.04<z_{\rm st}<0.08$, i.e., $200~{\rm Myr}\lesssim \Delta T_{\rm dur} \lesssim 800{\rm Myr}$, we hardly find a Coma-like cluster in our simulation samples made by our stochastic merger tree.
\par

For the above reason, in this section, we do not use a sample cluster in the merger tree but assume the merger history of a cluster with $M_{500}\approx 8\times10^{14}~M_\odot$, $z_{\rm obs} = 0.023$, and $z_{\rm st} = 0.06$.

We adopt the fiducial mass ratio of $\xi = 0.6$.
Prior to the merger event, the cluster mass evolves through mass accretion at a rate consistent with that assumed for the merger trees (Sect.~\ref{sec:merger_tree}).
The injection of CR is followed for $0<z<3$ using the model of Sect.~\ref{sec:injection}.
For an accurate modeling, we adopt the observed profiles of the thermal electron density \citep[][]{Briel_1992} and the temperature \citep[][]{Reiprich_Bohringer_2002}.
The profiles of the magnetic field and the turbulence are modeled with Eqs.~(\ref{eq:B_profile}) and (\ref{eq:turb_prof}), respectively.
\par


\begin{figure*}
    \centering
    \includegraphics[scale = 0.8]{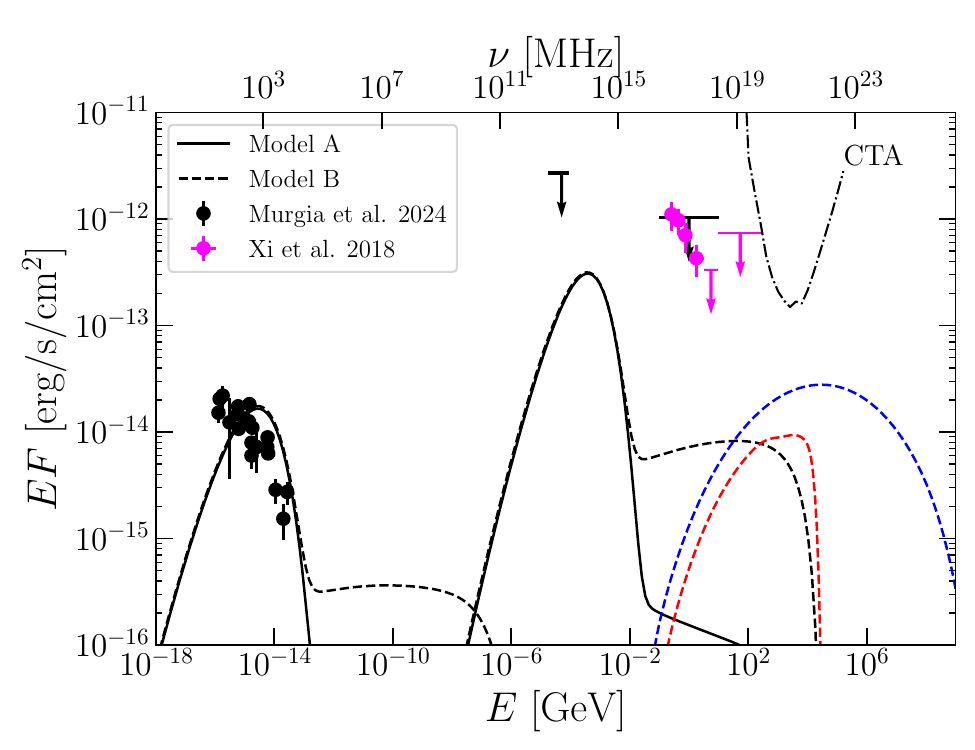}
    \plottwo{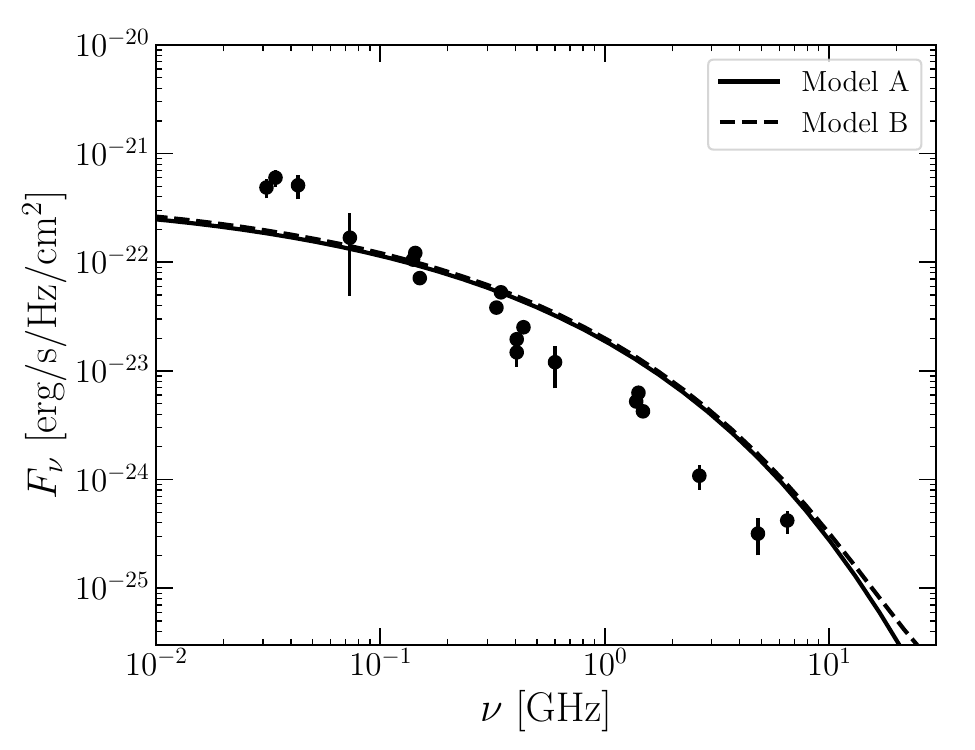}{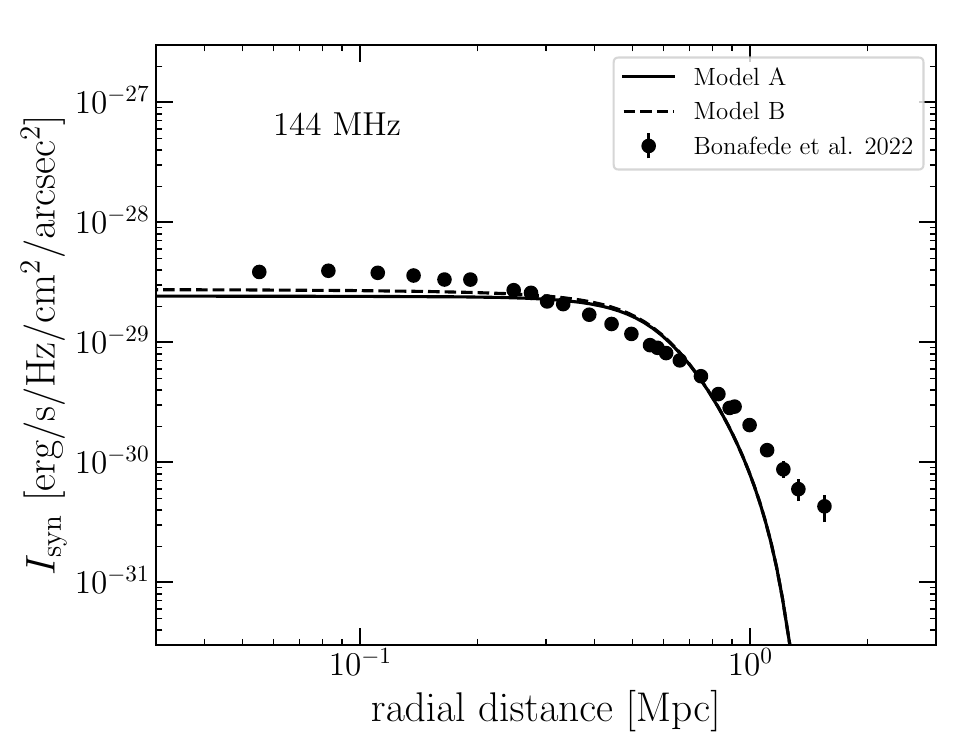}
    \caption{
    Upper panel: Spectra of non-thermal emissions of the Coma cluster. 
    The synchrotron and the inverse-Compton emissions of CREs (black), hadronic gamma-ray (red), and neutrino (blue) emissions of CRPs are shown for Model A (solid) and B (dashed).
    To compare with the radio observation, we assume the aperture radius of $r_{\rm ap} = 500$ kpc \citep[e.g.,][]{Brunetti_2013}. The data points are taken from \citet{Murgia_2024} for the radio emission and \citet{Xi_2018} for the possible detection of the gamma-ray with Fermi. The black bars with arrows show the upper limit on the non-thermal X-ray \citep[][]{Wik_2011} and the GeV gamma-ray \citep[][]{Ackermann_2016_Coma}. The black dotted line shows the sensitivity of CTA.
    Lower left panel: A zoomed plot of the synchrotron spectra.
    Lower right panel:  Synchrotron brightness profile of the Coma cluster at 144 MHz. The horizontal axis is the radial distance in the physical coordinate. The data points are taken from \citet{Bonafede_2022}.
    }
    \label{fig:Coma}
\end{figure*}

In our current model, we find that $L_e^{\rm inj} \sim 3.0\times10^{40}$ erg/s is compatible with the observed radio flux. 
In the upper panel of Fig.~\ref{fig:Coma}, we show the multi-wavelength spectra of Coma in our models.
The zoom-in plot for the radio spectrum is shown in the bottom left panel.
In Model B (dashed line), we assume $L_p^{\rm inj} = 3.0\times10^{40}$ erg/s, i.e., $L_p^{\rm inj}/L_e^{\rm inj} = 1$. 
In this case, the synchrotron emission of the secondary CREs is negligible compared to that of the primary CREs below 10 GHz.
A CRP injection rate approximately ten times larger than that of our Model B, in which secondary and primary electrons contribute comparably to GHz emission, does not appear to be in conflict with the Coma spectrum.
However, in the following sections, we show that such a model would overproduce the number of detectable RHs. An even larger $L_{p}^{\rm inj}$, such as $L_{p}^{\rm inj} \gtrsim 10^{42}$ erg/s, would be incompatible with the observation of the high-energy neutrino background (Sect.~\ref{sec:hadron}).
Our models roughly reproduce the extension of the Coma RH, $r \approx 500$ kpc (lower right panel).
\par

Our models are in a good agreement with the observation of the Coma RH.
Nevertheless, there are some discrepancies between the data points and the model predictions in Fig.~\ref{fig:Coma}.
For example, the radio spectra are slightly harder than the observed one, and the models underpredict the 144 MHz brightness in $r > 1$ Mpc.
Improvements to the fit to the data points may be achieved by tuning the model parameters for the detailed points, such as the cutoff shapes of the diffusion coefficient or the radial CR injection profile.
However, in this study, we do not proceed to a systematic search of optimal parameters for Coma, as such fine-tuned parameters may not be applicable to other clusters.
Therefore, we emphasize that the constraints on $L_e^{\rm inj}$ or $L_p^{\rm inj}$ reported in this study should be regarded as order-of-magnitude estimates.

\par


\begin{table}[]
    \centering
    \caption{Comparison of the five models considered in this work. Model A and Model B are our fiducial models. Model B and Model D are different in the spatial diffusion. \label{tab:models}}
    \begin{tabular}{ccccc}
    \hline
         models & $L_e^{\rm inj}$ [erg/s] & $L_p^{\rm inj}$ [erg/s] & $p_{0,e}/(m_e c)$ & $f_{\rm comp}^{\rm pre}$ \\
         \hline
         A  & $3.0\times10^{40}$ & 0 & 300 & 0.3 \\ 
         B  & $3.0\times10^{40}$ & $3.0\times10^{40}$ & 300 & 0.3 \\ 
         C & $5.0\times10^{40}$ & $1.5\times10^{41}$ & 300 & 0 \\ 
         D &$ 3.0 \times10^{40}$ & $1.0\times10^{41}$ & 300 & 0.3 \\ 
         E & $1.5\times10^{41}$ & 0 & 1 & 0.3 \\ 
         \hline
    \end{tabular}
\end{table}


The predicted non-thermal X-ray and gamma-ray fluxes are consistent with the upper limits \citep[][]{Wik_2011,Ackermann_2016_Coma}.
The {\it Fermi} observation of gamma rays from Coma is still controversial. 
While several studies have claimed the detection \citep[][]{Xi_2018,Abdollahi_2020,Ballet_2020,Adam_2021}, it was not reported in \citet{Ackermann_2016_Coma}.
Our model flux of gamma-rays is far below the gamma-ray flux reported in \citet{Xi_2018}.
The conventional secondary-electron model \citep[e.g.,][]{Dennison_1980ApJ...239L..93D,Blasi_Colafrancesco_1999,Kushnir_2024} or the secondary-electron reacceleration models \citep[][]{Adam_2021,paperI} is more favorable to explain the gamma-ray detection.
However, considering the RH statistics, the adopted parameters in those models may be optimistic as discussed in \citet{paperII}.
\par

In summary, we find that the normalization of the injection rate of primary CREs should be $L_{e}^{\rm inj} \sim 3\times10^{40}$ erg/s in order to explain the Coma flux.
The contribution of secondary electrons to the observed emission is negligible when $L_{p}^{\rm inj} \lesssim 3\times10^{41}$ erg/s.
However, one should note that these values are derived under the assumption of the internal-source injection model explained in Sect.~\ref{sec:injection}.
Furthermore, the production rate of the secondary electrons is affected by the reacceleration in the pre-merger state (see Sect.~\ref{sec:modelC}).
\par

In the ``primary-dominant model" presented in \citet{paperI}, $L_{p}^{\rm inj} \sim 10^{43}$ erg/s is compatible with the Coma spectrum. The discrepancy by a few orders of magnitude is due to the following assumptions in our model: (1) The present study exclusively considers scenarios where the primary electrons dominate over the secondary electrons in terms of radio emission up to 10 GHz. (2) The ICM evolution is modeled over a timescale exceeding 10 Gyr, spanning $0.02<z<3$, whereas \citet{paperI} assumed a fixed ICM profile and solved FP equations for only 4 Gyr (approximately $0.02<z<0.45$). (3) The CR injection profile is assumed to be proportional to the galaxy density profile (i.e., the internal-source model), whereas \citet{paperI} assumed an injection profile extended to large radii (see their Fig.~5). (4) The gentle reacceleration in the pre-merger state is included.
Approximately, the assumption (1) reduces the injection rate by a factor of ten, while the others each reduce it by a factor of three.
Those updates for the CR injection and acceleration mentioned above are based on more reasonable assumptions than those in our previous study.
\par


\section{Statistical properties of RHs}\label{sec:stat_RH}
The method developed in this study can be used to simulate the formation and decay of RHs during the formation history of GCs. In this section we discuss the statistical properties of RHs, such as their abundance, redshift distribution, and mass-radio power relation.
We compare our results with the recent LoTSS-DR2 survey at 140 MHz \citep[][]{Botteon_2022_LoTSS}.
\par

To reduce the computational cost, we use 400 trees whose $M_{z0}$ are in a limited range of $[10^{14}~M_\odot,10^{16}~M_\odot]$. The consistency to the Tinker mass function is ensured by multiplying a factor of $2000/400$ to $\Delta(\log_{10}M_{z0,i})$ (Eq.~(\ref{eq:tree_weight})). 
By using only 200 trees in the same mass range, we obtain the results qualitatively similar to those presented in this section. 
\par

We make a mock catalog of X-ray or SZ-selected GCs before considering the radio detectability.
We randomly pick a cluster at a given redshift $z$ from the FP-simulated halos using the probability of Eq.~(\ref{eq:MT_prob}).
We adopt sufficiently small $\Delta z$ so that $P_i\Delta z$ is guaranteed to be less than 1.
Using the selection criteria of the X-ray or SZ survey, we determine whether the GC is listed in the catalog or not.
When we model the LoTSS-DR2 survey, we use the 50\% Planck completeness line in the $M_{500}-z$ plane as the selection criterion, following \citet{Cassano_2023}.
We iterate the above selection until the number of GCs in the mock catalog reaches {\rm 164}, which is the number of Planck-SZ GCs above 50\% Planck completeness line in the survey \citep[][]{Cassano_2023}.
The LoTSS-DR2 survey studied the GCs in the region of 27\% of the northern sky, and the redshift of the clusters ranges from $0.07<z<0.5$. 
\par

We obtain the synchrotron flux of each GC in the mock catalog using the CR distribution calculated with the FP equations, and compare it with the sensitivity of the radio observation.
The sensitivity of the LoTSS-DR2 survey is modeled as follows \citep[][]{Cassano_2023}:
\begin{align}\label{eq:f_min_2}
	f_{\rm min} &= 4.44\times10^{-3}~[{\rm mJy}]\nonumber \\
    &~\times \zeta \left(\frac{F_{\rm rms}}{10~\mu{\rm Jy}}\right)\left(\frac{10~{\rm arcsec}}{\theta_{\rm b}}\right)\left(\frac{\theta_{\rm H}}{{\rm arcsec}}\right),
\end{align}
where $\zeta = 5$, and $F_{\rm rms} = 200~\mu$Jy/beam. The beam size $\theta_{\rm b} = \theta_{\rm b}(z)$ depends on the redshift with a fixed linear size of 150 kpc.
We calculate the flux within $r<3r_{\rm e}$, where $r_{\rm e} = 170$ kpc. 
\par

\begin{figure*}
    \centering
    \plottwo{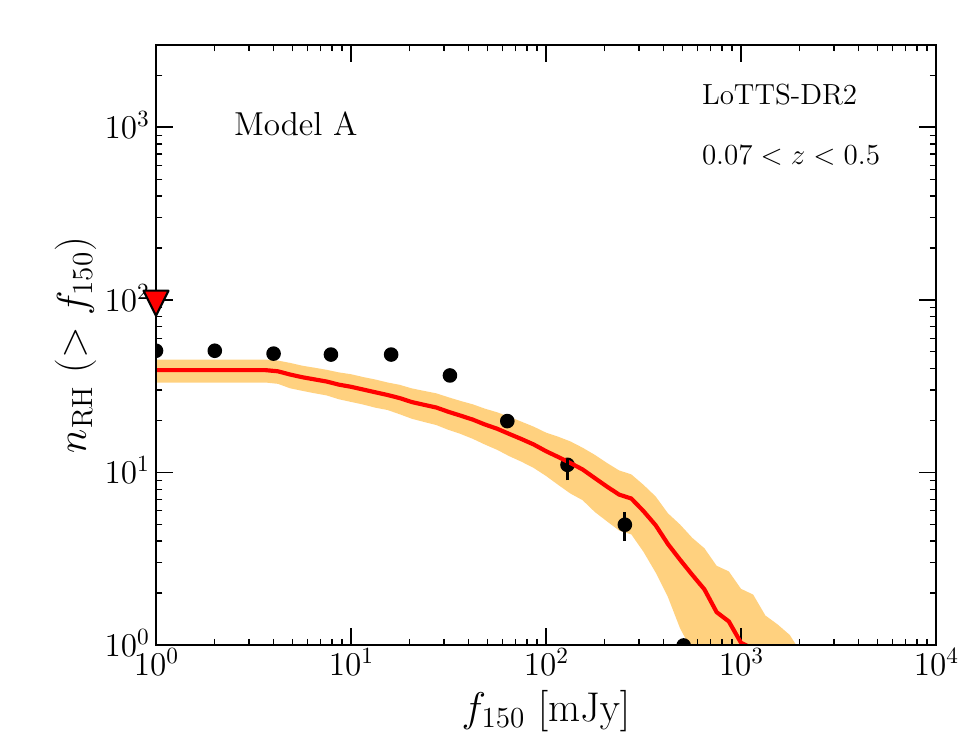}{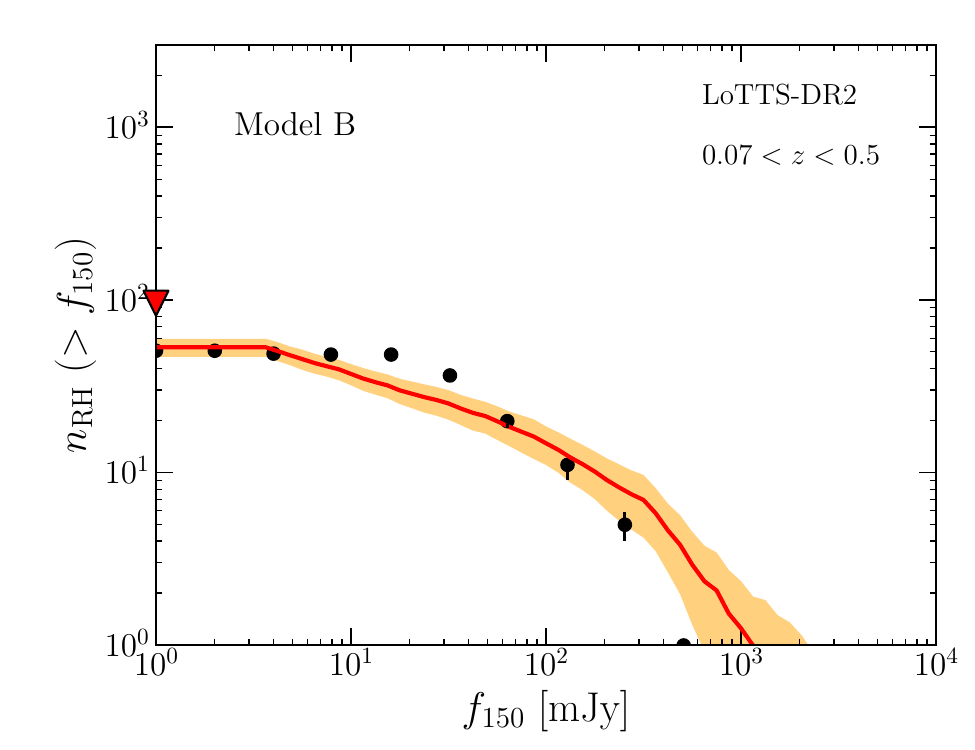}
    \plottwo{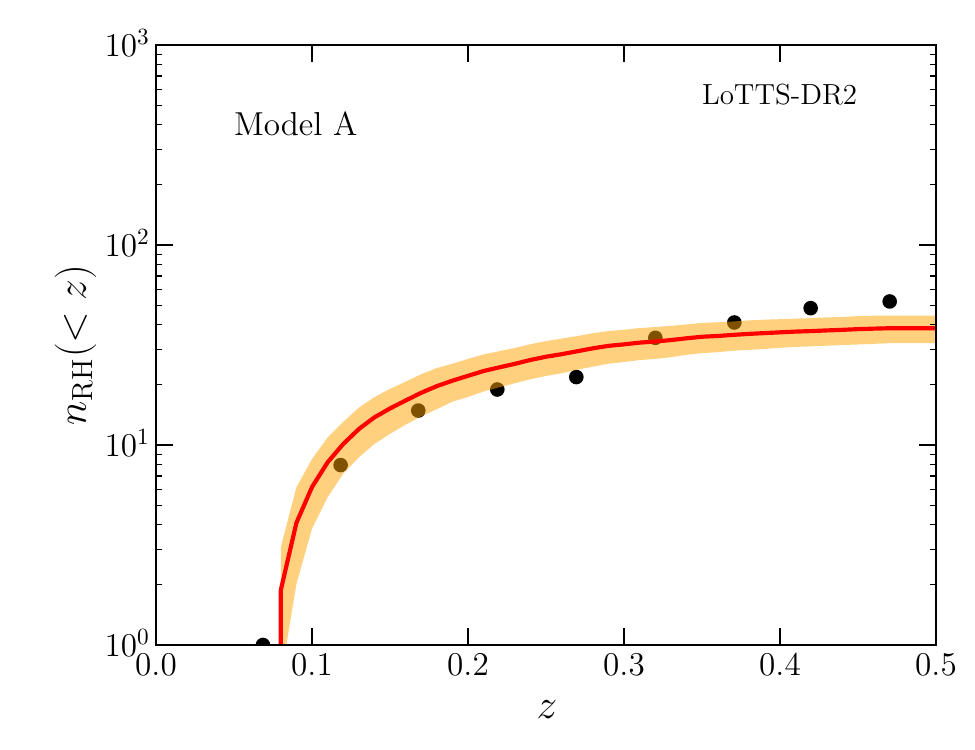}{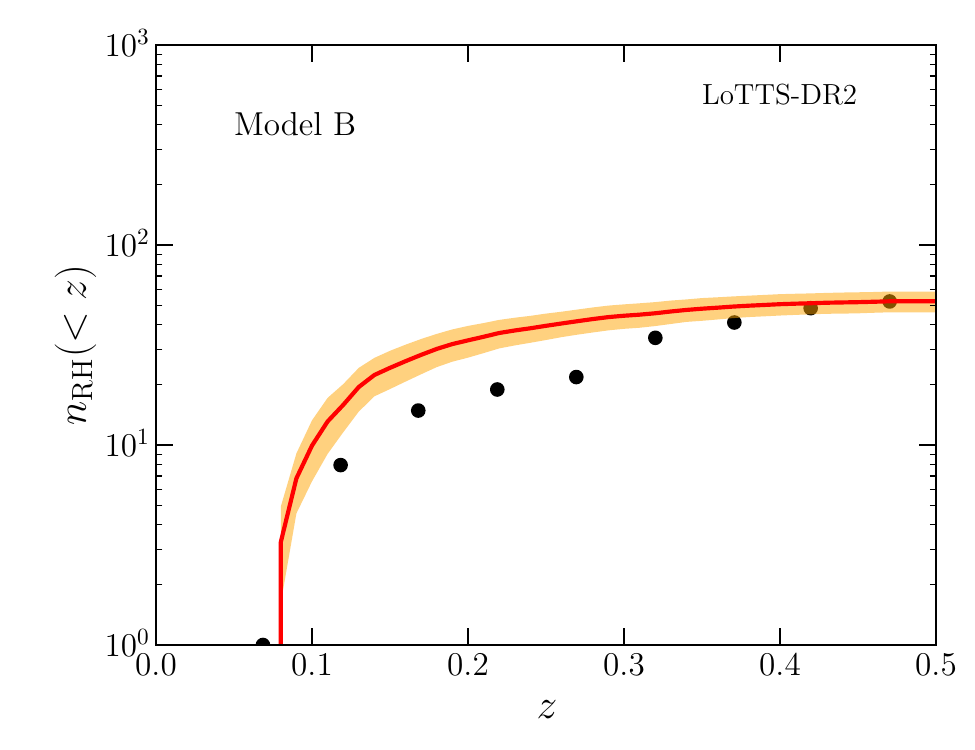}
    \caption{
    Upper panels: cumulative number count of RHs versus observed radio flux at 150 MHz. Our model curves are based on the sensitivity and the sky coverage of the LoTSS-DR2 survey. The red line shows the mean of the $N = 100$  mock catalogs of {\rm 164} samples, and the orange band shows 1$\sigma$ error. The right and left panels show the results for models A and B, respectively.
    The red arrow shows the upper limit of the cumulative number count discussed in \citet{Cassano_2023}.
    Lower panels: cumulative number of RHs as a function of redshift.
    }\label{fig:nRH}
\end{figure*}

Fig.~\ref{fig:nRH} compares the cumulative number counts of RHs in the LoTSS survey and our simulation. 
We perform the MC sampling of $N = 100$  mock catalogs.
Both models A and B are generally in good agreement with the observed number count (data points).
Our results are consistent with those of \citet{Cassano_2016}, i.e., the occurrence of RHs can be explained with $\xi_{\rm th}\approx0.2$ and $T_{\rm dur}\approx 3$ Gyr.
In Appendix~\ref{app:T_dur}, we explore the case of a shorter $T_{\rm dur}$ ($\approx 1.5$ Gyr).
By adopting a larger $L^{\rm inj}_e$ ($\approx 1.5\times10^{41}$ erg/s), that model can also explain the number of RHs, but the fit to the flux distribution becomes worth than Model A or B.

\par
%

The emission from secondary CREs is small compared to the primary CREs under the assumption of $L_p^{\rm inj}/L_e^{\rm inj}\sim1$ (Model B, see Tab.~\ref{tab:models}).
However, in the bottom right panel of Fig.~\ref{fig:nRH}, Model B overpredicts the cumulative number count in $z<0.3$, while Model A (bottom right), where $L_p^{\rm inj}=0$, matches well with the observation. 
As discussed in \citet{paperII}, the secondary production rate does not decay even after the end of the reacceleration due to the inefficient cooling of CRPs. Consequently, the effect of secondaries becomes more apparent in lower redshifts.

\par

Solving the FP equation to follow the injection and reacceleration of CRPs over a cosmological timescale, we have obtained a much tighter constraint on  $L_p^{\rm inj}$ (e.g., $<10^{41}$ erg/s) than those reported in the previous studies.
In \citet{paperII}, where $T_{\rm dur}=t_{\rm eddy}\approx700$ Myr, the injection rate can be as large as $L_p\sim 10^{43}$ erg/s. However, an unrealistically small value for $\xi_{\rm th}$ is required with this short $T_{\rm dur}$ to reproduce the RH number counts. In the current study, using a more reasonable set of parameters, we find that the Gyr-lasting reacceleration of CRPs has a great impact on the production rate of secondaries.
\par


\begin{figure*}
    \centering
    \centering
    \begin{minipage}[b]{0.32\textwidth}
    \centering
    \includegraphics[scale=0.37]{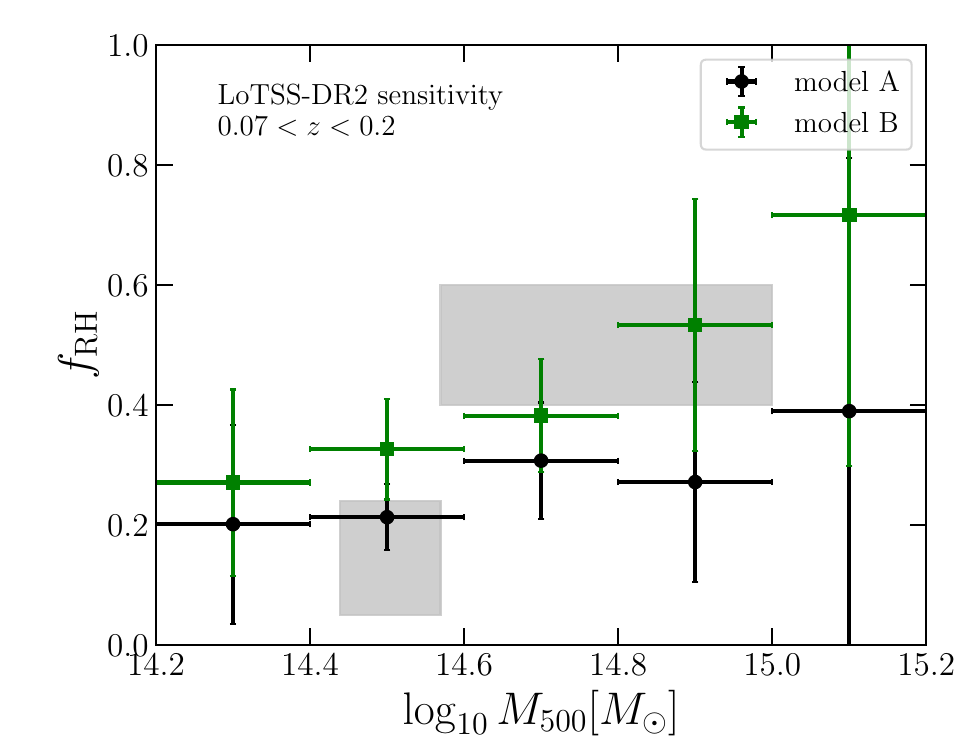}
    \end{minipage}
    \begin{minipage}[b]{0.32\textwidth}
    \centering
    \includegraphics[scale=0.37]{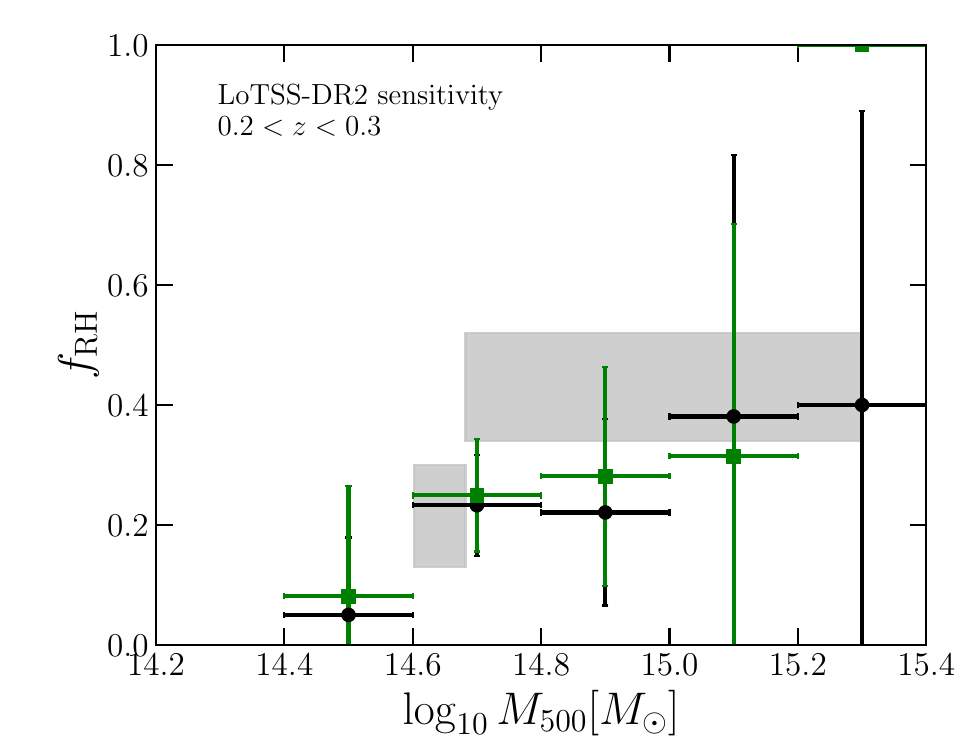}
    \end{minipage}
    \begin{minipage}[b]{0.32\textwidth}
    \centering
    \includegraphics[scale=0.37]{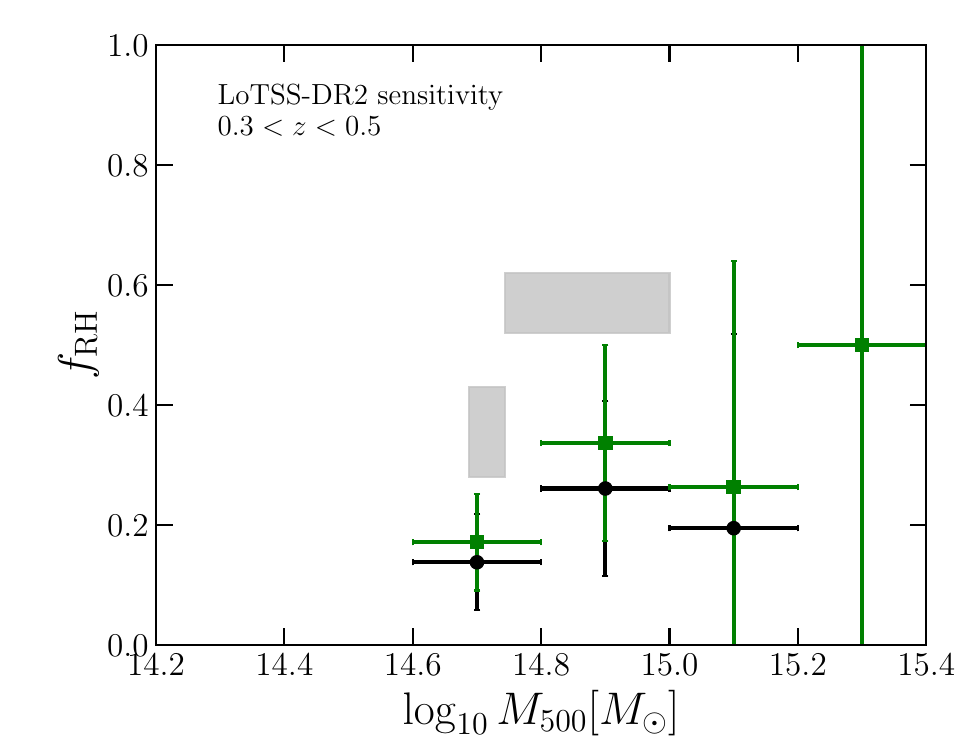}
    \end{minipage}
    \caption{
    The fraction of RHs versus cluster mass in the LoTSS-DR2 survey. 
    The data points and the error bars show the mean values of $f_{\rm RH}$ for $N = 100$ mock catalogs of 164 samples and 1$\sigma$ error, respectively.
    The shaded regions show the observed fraction including the uncertainty associated with the error on the mass estimate, adopted from \citet{Cassano_2023}.
    }\label{fig:fRH}
\end{figure*}

Fig.~\ref{fig:fRH} shows the fraction of RHs, $f_{\rm RH}$, as a function of cluster mass.
The data points with error bars and the shaded regions show our MC simulation and the observation \citep[][]{Cassano_2023}, respectively.
As previously observed by \citet{Cuciti_2015,Cuciti_2021b} and \citet{Cassano_2023}, $f_{\rm RH}$ increases with the cluster mass.
This trend is naturally expected in the TTD reacceleration model, where $D_{pp} \propto M^{1/3}$ (Eq.~(\ref{eq:tacc_scale})).
Our models can explain the typical value of $f_{\rm RH}\approx0.4$ and the mass trend.
The models tend to underpredict $f_{\rm RH}$ in the highest redshift bin ($z > 0.3$), although the discrepancy is not robust due to the large error.
A steeper redshift evolution of the CR injection rate or more efficient reacceleration at high redshift than our model may be favorable.
It should be noted that the predicted $f_{\rm RH}$-$M_{500}$ relation in the redshift range of $0.3<z<0.5$ becomes comparable to the observed one if $M_{500}$ is shifted by only a factor of $\approx2.5$.

\par

\begin{figure*}
    \centering
    \plottwo{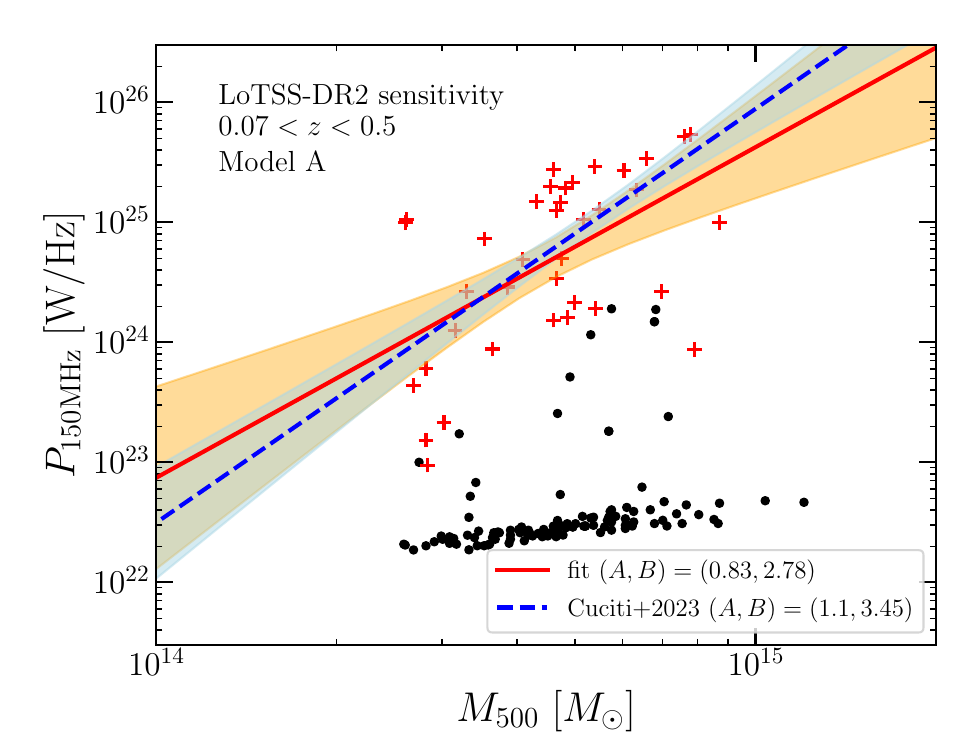}{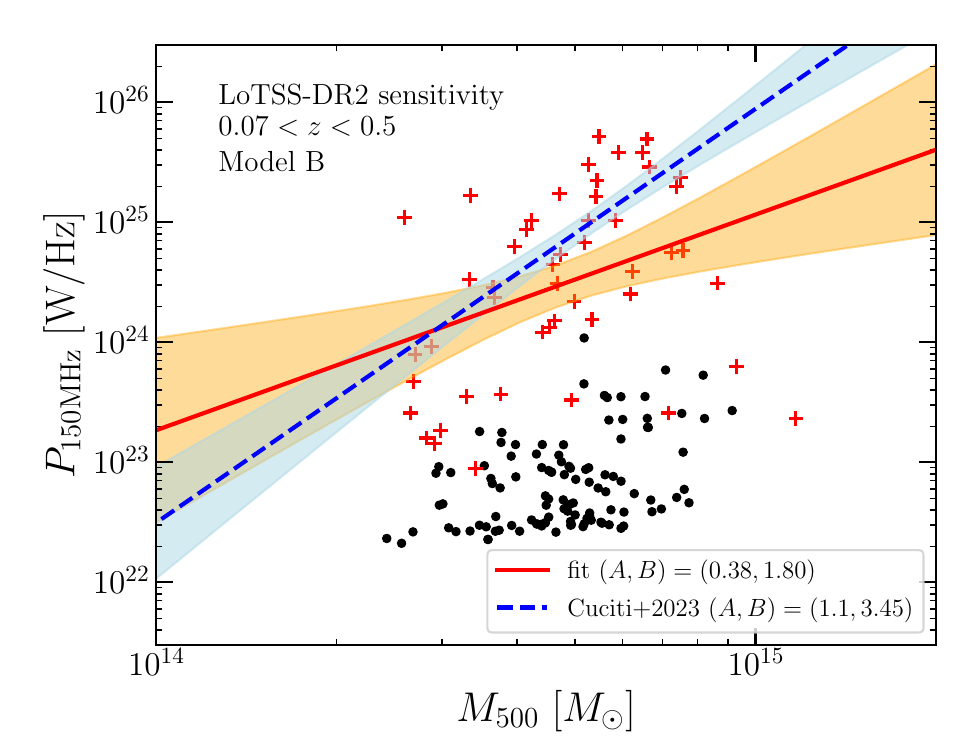}
    \caption{
    Mass -- radio power relation at 150 MHz. The red points correspond to the GCs in the mock catalog detectable with the LoTSS sensitivity, while the black points show the undetectable GCs. 
    The detectability of each cluster is determined by comparing the radio flux with the sensitivity of Eq.~(\ref{eq:f_min_2}).
    The red lines and the orange regions show the least mean square fit for the detectable GCs (red points) and its 95\% confidence interval, respectively.
    The blue dashed line and light-blue region correspond to the observed relation reported in \citet{Cuciti_2023}.
    }\label{fig:P-M}
\end{figure*}

The mass--radio power relation at 150 MHz {\rm in a mock sample} is shown Fig.~\ref{fig:P-M}.
The red points represent the RHs whose radio flux integrated within $r<3r_{\rm e}$ is larger than the LoTSS-DR2 sensitivity (Eq.~(\ref{eq:f_min_2})), while the black points represent GCs whose radio flux is below the detection limit.
Considering the redshift of each cluster, we convert the radio flux into the radio power per frequency $P_{\rm 150MHz}$.
Note that some detectable RHs display a radio power smaller than the typical power of non-RHs and vice versa.
This happens when a dim (bright) cluster is located at a low (high) redshift. 

\par

The bimodality of RHs and non-RH GCs is clearly seen in our Model A.
The typical luminosity of non-RH GCs is as low as $P_{\rm 150 MHz} \sim 10^{22}$ W/Hz, which is powered by the continuous injection of primary CREs. In contrast, RHs have $P_{\rm 150 MHz} \sim 10^{24.5}$ W/Hz. 
The probability of finding a GC with an intermediate luminosity, $10^{23}~{\rm W/Hz} < P_{\rm 150 MHz} < 10^{24}~{\rm W/Hz}$, is low due to the rapid increase and subsequent decay in radio luminosity seen in the primary-electron reacceleration model \citep[e.g.,][]{Brunetti_2009,paperII}.
The timescale of the rise and decay is comparable to the reacceleration timescale $t_{\rm acc}\approx 500$ Myr.

\par

The mass--radio power relation  for RHs (red points) is fitted with the following power-law function:
\begin{equation}
    \log\left(\frac{P_{\rm 150MHz}}{10^{24.5}{\rm W/Hz}}\right) = B\log\left(\frac{M_{500}}{10^{14.9}M_\odot}\right)+A.
\end{equation}

We adopt the least mean square method and the resulting curve is shown with the red line in Fig.~\ref{fig:P-M}.
The orange region shows the 95\% confidence interval of the fitting.
We confirm that the conventional reacceleration model of primary electrons, i.e., Model A, can reproduce the steep mass--power relation.
On the other hand, the predicted mass-power relation is less steep in Model B than in Model A due to the increased number of the RHs with intermediate luminosities.
Those intermediate RHs are powered by the secondary CRE production through the $pp$ collisions.
This result is consistent with Fig.~\ref{fig:nRH}, which demonstrates an increased number of low-z RHs.
Due to the secondary CRE production by the re-accelerated CRPs, whose ``lifetime" is much longer than the cooling time of CREs, the GCs with similar masses exhibit varying radio luminosities depending on their merger history.
A low CRP injection rate like Model A is preferred in reproducing the RH number count and the mass–power relation under the assumptions of the internal-source injection and $T_{\rm dur}\approx$ 3 Gyr. The contribution of secondary CREs to the observed radio emission should be marginal.
\par

\begin{figure*}
    \centering
    \plottwo{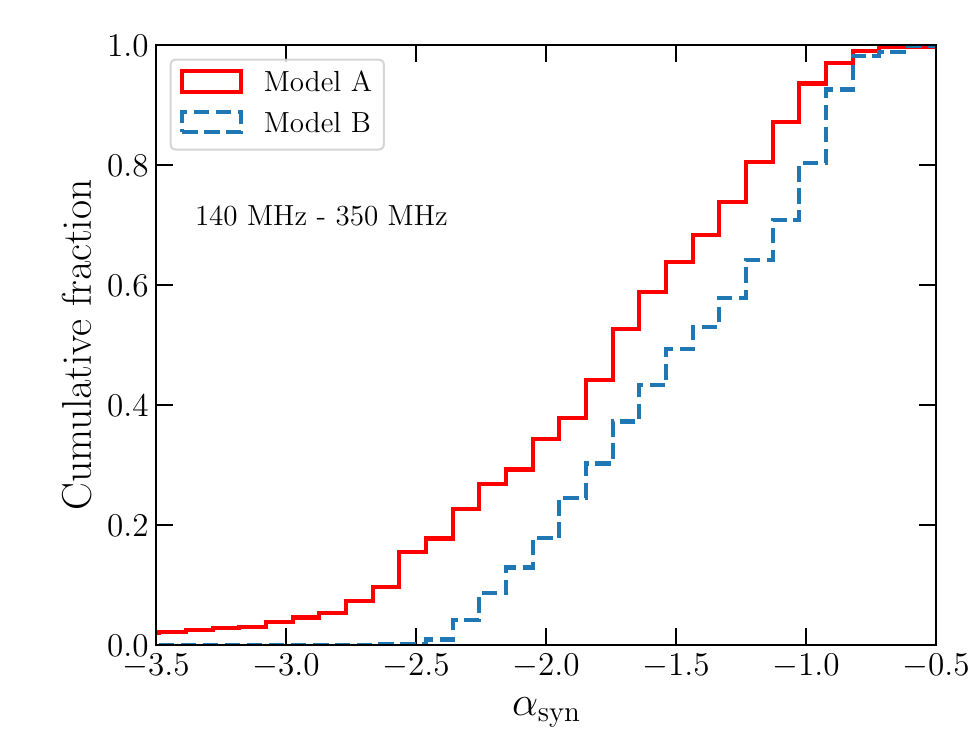}{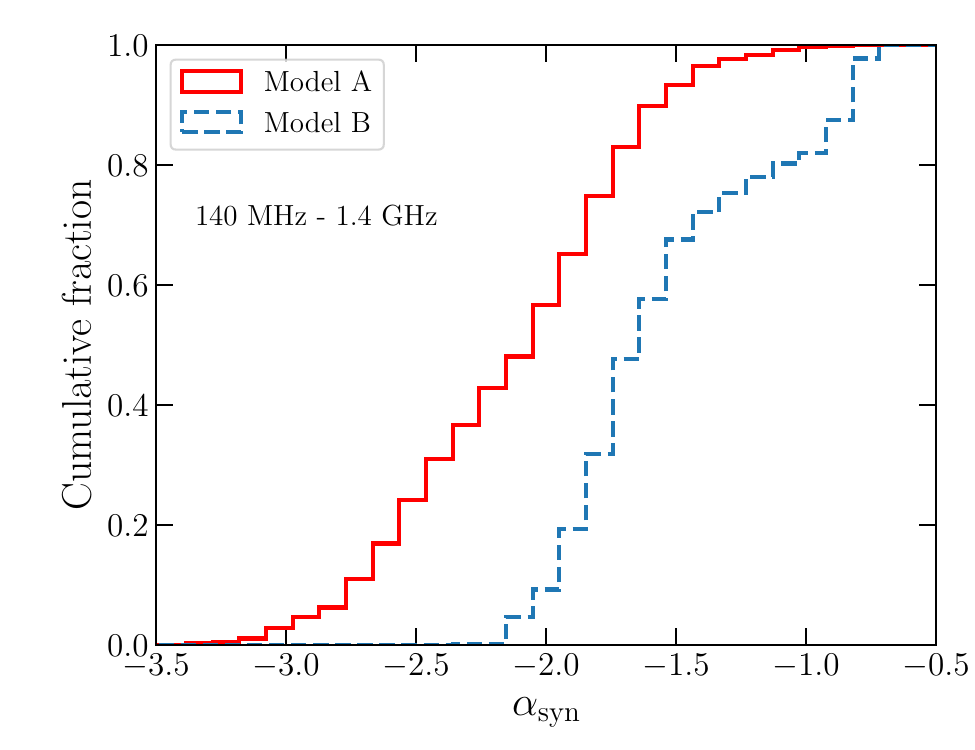}
    \caption{The cumulative fraction of RHs versus radio spectral index. The solid and dashed lines represent models A and B, respectively. The right and left panels show the results for the indices measured in 140 MHz$<\nu<$350 MHz and 140 MHz$<\nu<$1.4 GHz, respectively. }
    \label{fig:spix}
\end{figure*}
The reacceleration model predicts that the inefficient acceleration in a merger with a small mass ratio $\xi$ results in a RH with a very steep radio spectrum \citep[e.g.,][]{Brunetti_2008,Cassano_2010_LOFAR,Brunetti_Jones_review}.
In the left panel of Fig.~\ref{fig:spix}, we show the cumulative fraction of RHs as a function of the radio spectral index between 140 MHz and 350 MHz in models A (solid) and B (dashed).
We find that more than half of the RHs have very steep spectral index, $\alpha_{\rm syn} < -1.5$, in this frequency range.
A similar expectation was obtained from the theoretical model in \citet{Cassano_2023}.
The fraction is slightly reduced in Model B due to the injection of secondary electrons with a hard energy spectrum.
The difference between the models is pronounced when the spectral index is measured in a wider frequency range (right panel).
Unfortunately, those predictions cannot be immediately tested by the observation due to the limited statistics of RHs observed in multi-frequency. 
There is a hint for the large fraction of ultra-steep spectrum radio halos (USSRHs) obtained by recent follow-up observation of LoTSS-DR1 RHs using the low-frequency band ($\sim50$ MHz) of LOFAR \citep[][]{Pasini_2024}.
\par


\begin{figure*}
    \centering
    \plottwo{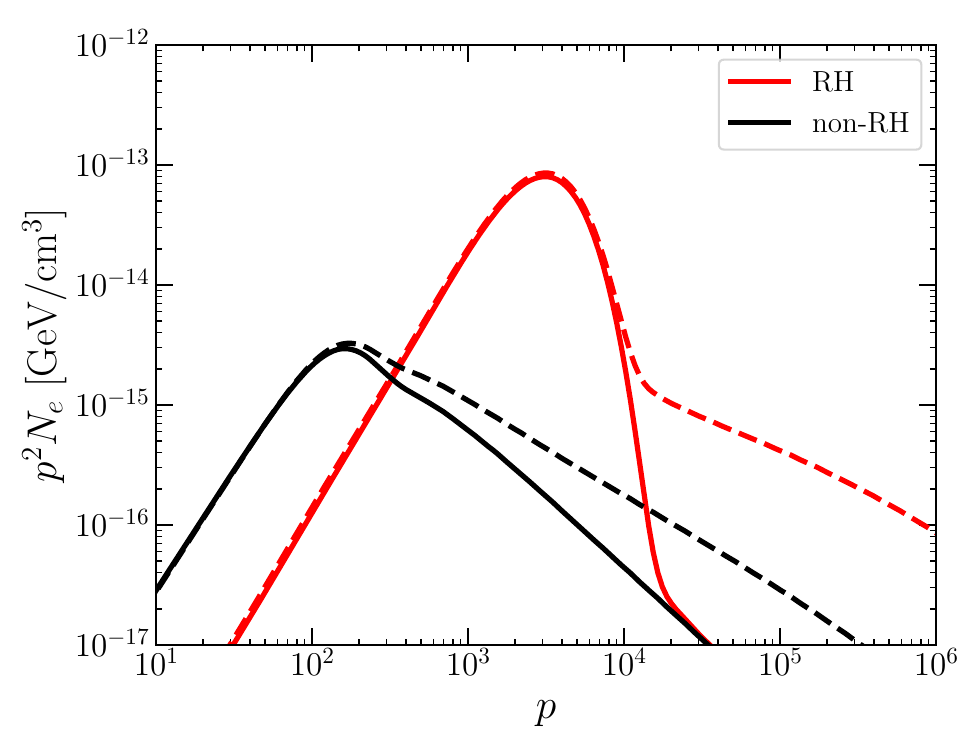}{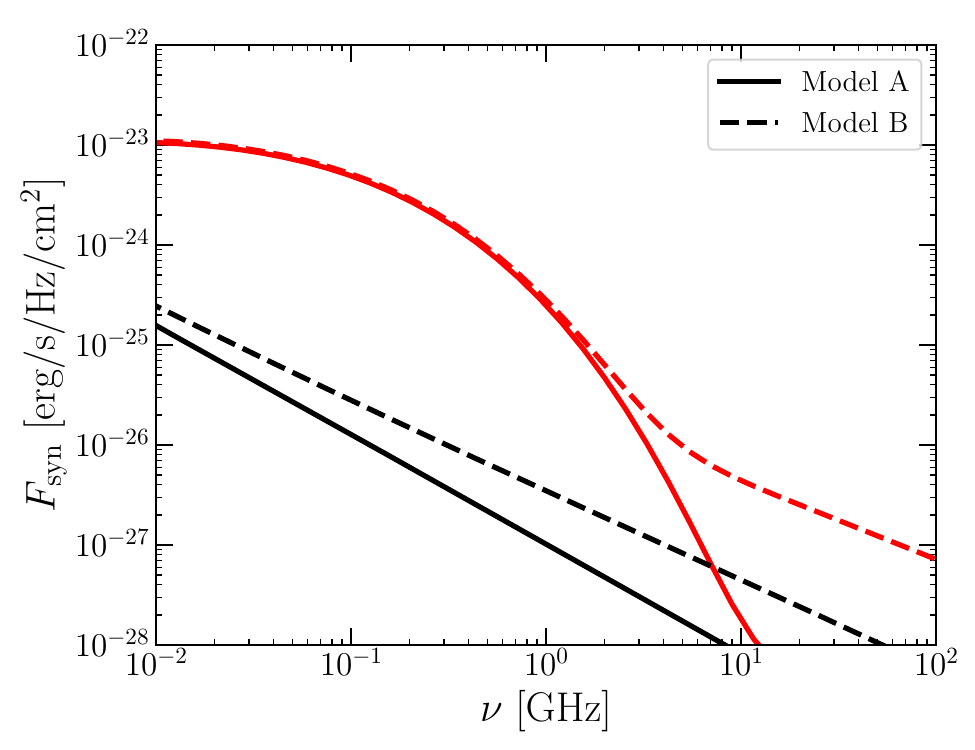}
    \caption{
    The CRE (left panel) and synchrotron (right panel) spectra in Model A (solid lines) and Model B (dashed lines). 
    The red and black lines show the spectra for the LoTSS-detectable and undetectable clusters, respectively. Both cluster have the mass of $M_{500} \approx 5\times10^{14}~M_\odot$ and located at $z = 0.15$.
    In the left panel, the horizontal axis shows the particle momentum normalized by $m_ec$.
    }
    \label{fig:CR_spec}
\end{figure*}

In Fig.~\ref{fig:CR_spec}, we show the spectra of CREs (left) and the synchrotron emission (right) for the examples of typical RH (red) and non-RH clusters (black) found in our simulation.
Both the clusters are located at $z = 0.15$ and have comparable masses, $M_{500} \approx 5\times 10^{14}~M_\odot$. The radio power of the example RH is typical of those observed for this mass, $P_{\rm 150MHz} =  1.2\times10^{25}$ W/Hz (see Fig.~\ref{fig:P-M}).
The non-RH cluster has $P_{\rm 150MHz} = 3.1\times10^{22}$ W/Hz in the model A.
The RH cluster underwent a merger with $\xi = 0.63$ at $z = 0.34$, and the reacceleration is still ongoing at $z = 0.15$.
On the other hand, the non-RH cluster experienced a merger with $\xi = 0.20$ at $z = 0.78$, and the reacceleration has ceased at $z = 0.40$.
In the RHs, the CREs are efficiently re-accelerated up to $p_e/(m_ec)\sim3\times10^3$, while the CRE spectrum shows a peak around $p_e/(m_ec)\sim 2\times10^2$ in the non-RH clusters.
The secondary-electron component in Model B appears above $p/(m_ec) > 10^4$.
The injection of those secondary CREs is enhanced by a factor of $\sim 20$ in the RHs due to the reacceleration of CRPs.
The excess of the synchrotron flux due to this component becomes significant above several GHz.
Thus, the statistical properties of RHs at 10 GHz would be important to constrain the CRP injection and reacceleration.
\par

\begin{figure*}
    \centering
    \plottwo{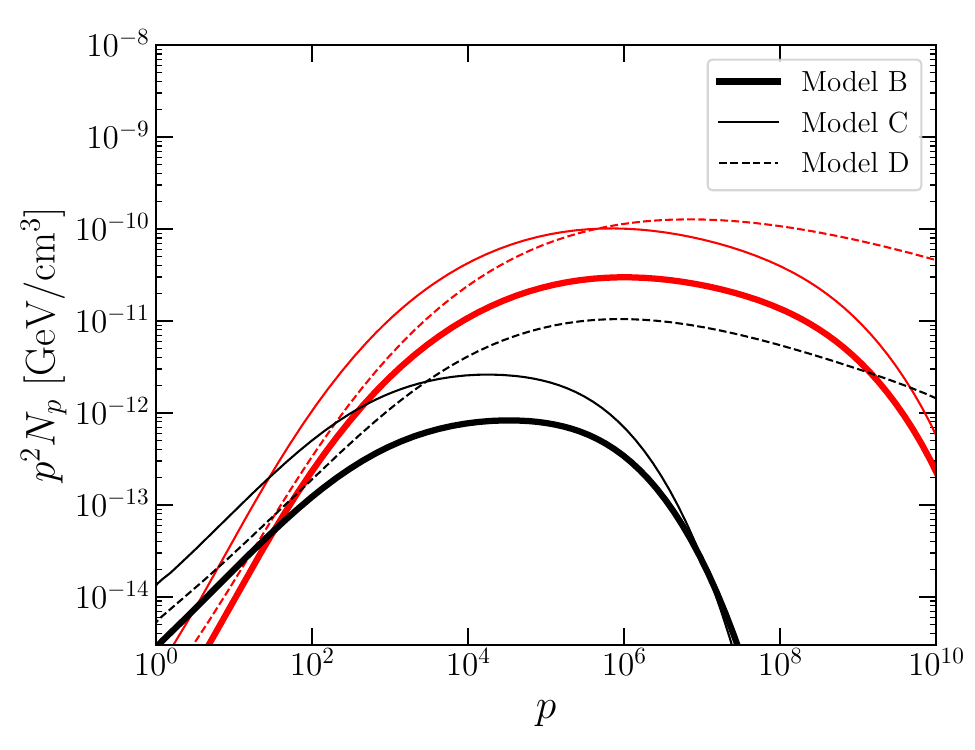}{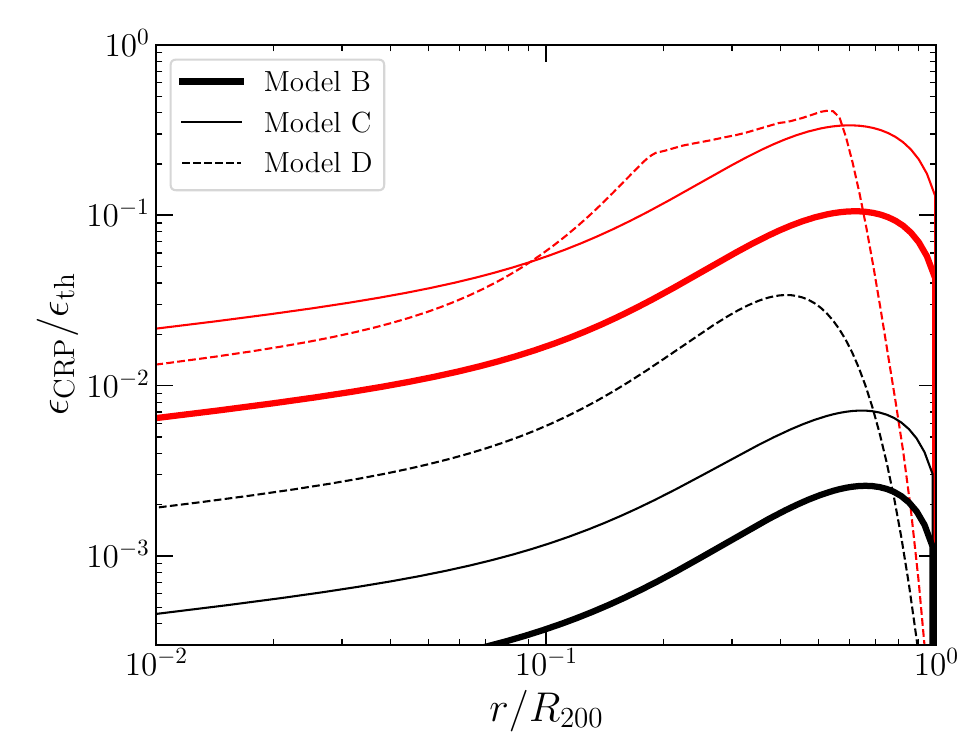}
    \caption{
    Left: CRP spectra for the typical RH (red) and non-RH (black) clusters.
    We consider the same clusters as in Fig.~\ref{fig:CR_spec}.
    The thick solid, thin solid, and thin dashed lines are for Model B, C, and D, respectively.
    Right: radial profile of the fraction of CRP energy density to the thermal energy density. 
    }
    \label{fig:CRP}
\end{figure*}

The left panel of Fig.~\ref{fig:CRP} shows the spectra of CRPs for the RH and non-RH GCs, as in Fig.~\ref{fig:CR_spec}.
The result for Model B is shown with the thick lines.
Notably, the CRP spectra for both clusters are clearly harder than the injection spectrum $N_p\propto p^{-2.2}$.
This hardening is attributed to the hard-sphere type acceleration. 
In contrast to CREs, the radiative loss of CRPs is inefficient throughout the entire energy range of interest ($1<p/(m_pc)<10^{10}$). In the timescale of $T_{\rm dur}\sim3$ Gyr, the diffusive escape is important for $p/(m_pc) > 10^6$. Consequently, the CRPs are efficiently accelerated up to $E_p^{\rm max} \sim 1-10$ PeV.
The typical duration of the reacceleration to explain the RH population ($T_{\rm dur} \sim 3$ Gyr) is much longer than $t_{\rm acc}$, which allows for a significant modification of the initial spectrum by the reacceleration.
In contrast, for the non-RH GCs, the less hard spectrum than the RHs is caused by the reacceleration associated with the mergers at high redshifts ($z > 0.5$) and the gentle reacceleration during the relaxed state.
This hard spectrum of CRPs in non-RH GCs is a remarkable discovery in this paper.
The expected hadronic emissions from those CRPs are discussed in detail in Sect.~\ref{sec:hadron}.
\par



The right panel of Fig.~\ref{fig:CRP} shows the radial profile of the ratio between the CRP energy density and the thermal energy density in Model B.
The results indicate that the ratio increases with radius, in accordance with the assumptions commonly made in RH models \citep[e.g.,][]{Brunetti_2017,paperI,Keshet_2023}.
In the RHs, the CRP energy density is $\sim1$\% of the thermal energy density in the central region and reaches to $\sim10$\% at large cluster radii ($r/R_{200} \sim 0.5 - 1$).
The energy density in the non-RH GCs is almost one order of magnitude smaller.
Note that the above results are derived under the assumption of the hard-sphere type acceleration. 


\section{High-energy emissions from CRPs}\label{sec:hadron}
As shown in Fig.~\ref{fig:CRP}, the energy spectrum of the re-accelerated CRPs becomes very hard.
In this section, we discuss the hadronic emissions from the $pp$ collisions between CRPs and the thermal protons.
Only Model B is discussed in this section.
\par


\begin{figure}
    \centering
    \plotone{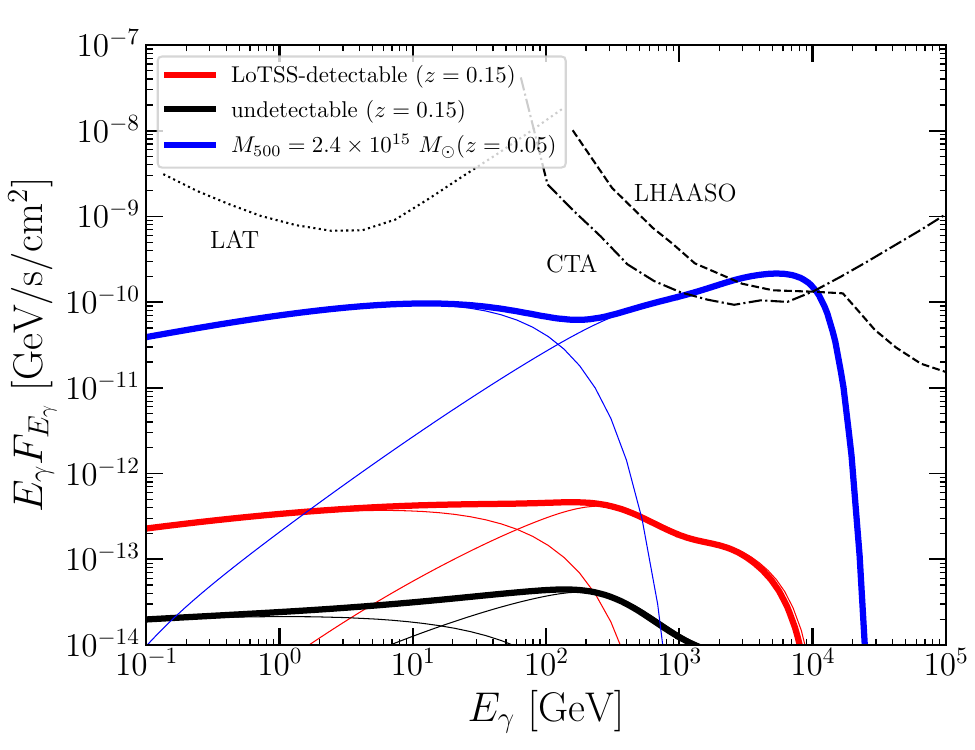}
    \caption{The gamma-ray spectra in Model B for the clusters in Fig.~\ref{fig:CR_spec} and \ref{fig:CRP}. The red and black lines correspond to the RH and non-RH GCs, respectively.
    Thin lines show the IC emission of CREs and the emission from the $\pi^0$ decay, while thick lines show the total flux. 
    The blue lines show the spectra for a very massive GC with $M_{500} = 2.4\times 10^{15}~M_\odot$ at $z = 0.05$.
    The sensitivities of Fermi-LAT, CTA, and LHASSO are also shown.}
    \label{fig:gamma}
\end{figure}

Fig.~\ref{fig:gamma} compares the gamma-ray spectra of a LoTSS-detectable GC and an undetectable GC, whose CR spectra are shown in Fig.~\ref{fig:CR_spec} and \ref{fig:CRP}. 
Both are located at $z = 0.15$.
In both cases, the gamma-ray above 100 GeV is dominated by the hadronic ($\pi^0\rightarrow 2\gamma$) component.
We find that the typical gamma-ray flux from the RHs is too dim to detect with Fermi, CTA, and LHASSO.
However, a detection of TeV gamma rays from a very massive GC with $M_{500} \approx 2.4\times 10^{15} M_\odot$ located at $z\leq 0.05$ might be attainable, as shown by the blue line.
This very bright GC in our sample experiences two consecutive major mergers at $z = 0.44$ with $\xi = 0.66$ and at $z = 0.18$ with $\xi = 0.54$. It achieves the radio power of $P_{\rm 150MHz} \approx 1.5\times 10^{26}$ W/Hz at $z = 0.05$.
However, the probability of finding such a GC is very low.
In our simulation, the expected number of GCs detectable with LHASSO sensitivity is $\approx1$ in the entire sky.
If a TeV gamma ray is detected from a GC without any GeV gamma-ray detection by Fermi, the CRP spectrum should be hard, as expected in our hard-sphere reacceleration model.
The Coma cluster may not produce detectable TeV-PeV gamma rays, given that the it is assumed to be at an early stage of the reacceleration process (see Sect~\ref{sec:Coma}).
\par

We calculate the isotropic background of the high-energy neutrinos produced by all GCs in the entire sky and compare it with the observation by IceCube.
The mean intensity of the background (in the unit of $dE/dE_\nu dt dAd\Omega$) can be calculated as follows \citep[e.g.,][]{paperIII}:
\begin{align}\label{eq:bkgd}
	E_\nu\Phi_\nu 
	 = \frac{c}{4\pi}\int dz\frac{1}{(1+z)H(z)}\int dL_{E_\nu}\frac{dN}{dL_{E_\nu} dV_{\rm c}}L_{E_\nu}, 
\end{align}
where $\frac{dN}{dL_{E_\nu} dV_{\rm c}}$ is the neutrino luminosity function, $L_{E_\nu}$ is the neutrino luminosity at $E_\nu$ in the unit of [GeV/s/GeV], and $D_{\rm L}(z)$ is the luminosity distance. 
In this calculation, we do not use the mock catalog of GCs generated via the MC sampling (Sect.~\ref{sec:stat_RH}).
Instead, we consider the neutrino luminosity of all of the GCs in the 400 merger trees at each redshift in order to calculate the neutrino luminosity function.
The number of GCs is calculated with Eq.~(\ref{eq:MT_prob}).
\par

\begin{figure}
    \centering
    \plotone{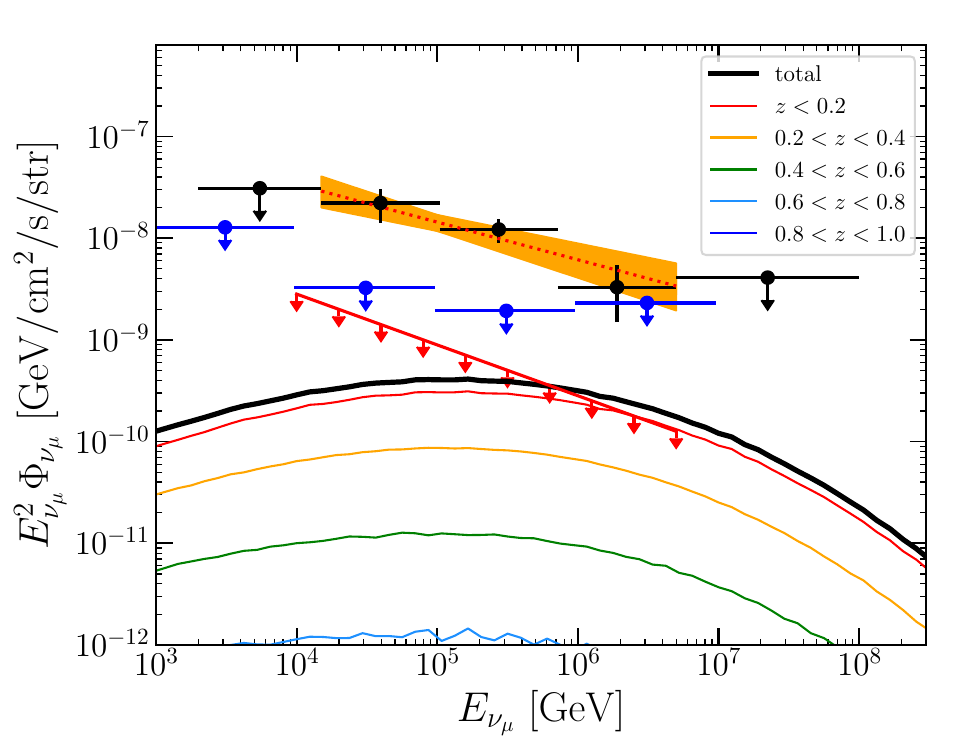}
    \caption{The spectrum of the isotropic neutrino background from GCs in Model B.
    The contributions from various redshift bin are shown with colored lines.
    The black points and the orange band show the all-sky neutrino intensity and the spectrum of IceCube track events \citep[][]{IceCube_2022_muontrack}.
    The red line with arrows shows the 90\% CL upper limit derived from the stacking analysis of the Planck-SZ GCs using the distance-weighting scheme \citep[][]{IceCube_2022_stack}. The blue arrows show the upper limits in quasi-differential energy bins.}
    \label{fig:neutrino}
\end{figure}

In Fig.~\ref{fig:neutrino}, we compare the predicted spectrum of the high-energy neutrino background with the upper limits reported by \citet{IceCube_2022_stack}.
The flux is smaller than our previous expectation in \citet{paperIII}, where we assumed that the secondary electrons dominate over the primary electrons.
The spectrum is harder than that in \citet{paperIII}, where the temporal evolution of the spectral shape was neglected for simplicity.

In our reacceleration model, the nearby GCs ($z < 0.2$) contribute the most to the total neutrino background. In contrast, the previous internal-source models without reacceleration predicted that the dominant contribution would come from $0.3\leq z \leq 1$ \citep[e.g.,][]{Fang_Olinto_2016,Hussain_2021}.
Our Model B is consistent with the neutrino upper limit derived for the differential energy bins (blue arrows), indicating that the nearby massive GCs cannot be the dominant source of the neutrino background.
Fig.~\ref{fig:neutrino} indicates that the neutrino limit excludes the CRP injection rate larger than $L_{\rm p}^{\rm inj} > 10^{42}$ erg/s.
\par

In summary, based on our model, in which the uncertain parameters are adjusted to be consistent with the statistics of the RH observations, cosmological structure formation history, and numerical simulations for turbulence in GCs, the detection of TeV gamma rays can be a challenging goal, while the neutrino limit given by the stacking analysis provides a meaningful constraint on the CR injection rate.
The distinctive aspect of our findings regarding hadronic emission in our model is the hard spectrum resulting from the reacceleration of CRPs. 
However, we emphasize that the above results are derived under the assumption of the conventional TTD acceleration ($D_{pp} \propto p^2$) and $T_{\rm dur}\approx$3 Gyr.
The results would be significantly altered in the case of a different reacceleration model, such as the Kolmogorov-type reacceleration $D_{pp}\propto p^{1/3}$ \citep[e.g.,][]{Fujita_2015}.
To explain the RH spectrum, the acceleration timescale of $\sim$GeV CRs should be similar for both the Kolmogorov and the hard-sphere models.
Because the acceleration of higher energy particles is suppressed in the Kolmogorov model, the resultant CRP spectrum above $\sim$10 GeV would be significantly softer than that reported in this work \citep[see also][]{paperI}.
In the Kolmogorov model, the detection of TeV gamma rays would be even more challenging, and the constraints from the IceCube limit would be much less stringent.
\par


\section{Discussion}\label{sec:discussion}
In the previous sections, we have obtained the injection rate of primary CRs in the fiducial reacceleration model.
The primary-electron reacceleration model is consistent with RH observations. While CRPs should also be present in the ICM, the CRP injection rate should be as low as $L_p \lesssim 10^{41}$ erg/s to avoid too many secondary electrons.
In the following section, we demonstrate that this upper limit is much smaller than the expectations in the previous studies (Sect.~\ref{sec:inj_problem}).
The low injection rate is required mainly because of the ``internal-source" assumption, where the injection rate decreases with the radial distance.
Although a larger injection rate is allowed when the CRPs are injected at large radii, as demonstrated in \citet{paperI}, the contribution from the internal sources seems unavoidable. 
The obtained upper limit for $L_p$ raises an important question of why the CRP injection is so low.
By examining the models with different assumptions, we study the generality of the issue.
It is found that the injection rate can be increased by a factor of $\approx3$ when the reacceleration in the pre-merger state is negligible (Model C, Sect.~\ref{sec:modelC}) or the spatial diffusion is as large as $D_{rr} \sim 10^{31}$ cm$^2$/s (Model D, Sect.~\ref{sec:modelD}), which can only slightly alleviate the problem.
Furthermore, we present other potential solutions, which should be investigated in future studies.

\subsection{Expected injection rates from various sources}\label{sec:inj_problem}
In this section, we compare the CR injection rate constrained in this study (Tab.~\ref{tab:models}) with the expected rate from the possible CR sources. 
Our findings indicate that that the former is smaller than the latter by a few orders of magnitude. 
This discrepancy may suggest the the inefficiency of CRP acceleration at those sources, or the necessity to revise the reacceleration model.
\par

\subsubsection{Supernovae in member galaxies}
The CRs can be accelerated at the shocks of supernova remnants (SNRs) in member galaxies and supplied to the ICM.
One can infer the rate at which CRs are injected into the ICM from the metallicity of the ICM, $Z_{\rm cl}$, as described in \citet{Volk_1996}. 
The typical value for $Z_{\rm cl}$ is found to be $Z_{\rm cl}\sim0.3Z_\odot$, where $Z_\odot\approx0.016$ is the solar metallicity.
This value exhibits little dependence on the cluster richness, redshift, and the dynamical state \citep[see][for review]{Mernier_Biffi_2022}.
The relatively large value and extended profile of $Z_{\rm cl}$ suggests that majority of the ICM metals are supplied at $z \geq 2$, i.e., the early-enrichment scenario \citep[e.g.,][]{Elbaz_1995,Fujita_2008,Loewenstein_2013,Biffi_2018,Blackwell_2022}.
The CR injection rate averaged over the Hubble time $t_{\rm H}$ can be estimated as
\begin{align}\label{eq:LCR_Fe}
    L_{p}^{\rm SB} &\lesssim \frac{N_{\rm SN}\eta_{\rm CR}E_{\rm SN}}{t_{\rm H}}, \nonumber\\  
    &\sim 10^{43.5}~{\rm erg/s}\left(\frac{N_{\rm SN}}{10^{11}}\right)\left(\frac{\eta_{\rm CR}}{0.1}\right)\left(\frac{t_{\rm H}}{13~{\rm Gyr}}\right)^{-1},
\end{align}
where $N_{\rm SN} = M_{\rm Fe}/\delta M_{Fe}$ is the total number of SNe required to explain the total Fe mass in the ICM, $M_{\rm Fe}$ \citep[][]{Volk_1996}, $\delta M_{\rm Fe}\sim 0.1M_\odot$ is the iron yield per SN \citep[e.g.,][]{Mernier_Biffi_2022}, $\eta_{\rm CR}\sim 0.1$ is the CR acceleration efficiency at the SNR shock, and $E_{\rm SN} \sim 10^{51}$ erg is the typical explosion energy of SNe. 
The total iron mass is related to the metallicity as $M_{\rm Fe} \approx 1.3\times 10^{10} M_\odot \left(\frac{Z_{\rm cl}}{0.3Z_\odot}\right)\left(\frac{M_{\rm gas}}{10^{13.5}~M_\odot}\right)$, where $M_{\rm gas} \approx f_{\rm b}M_{200}$ is the total baryonic mass of the GC, and we assume the abundance ratio in the solar neighborhood \citep[e.g.,][]{Lodders_2009}.
CRs in galaxies are considered to escape to intergalactic space via the diffusion process. In this case, a large CR injection rate like Eq.~(\ref{eq:LCR_Fe}) is inevitable. If the dominant CR transport mechanism is the advection with galactic winds, CRs coupled with expanding winds can be cooled via adiabatic cooling.
In the interstellar medium of starburst galaxies, the CRs can be severely cooled by the pp collision (Sect.~\ref{sec:CRPloss}).
\par

\subsubsection{AGN jets}
CRs can be provided from the relativistic jets driven by AGNs \citep[e.g.,][]{Hardcastle_Croston_2020}. The radio lobe surrounding the powerful jet found in the central region of GCs excavates the ICM. The kinetic energy estimated from the size of the X-ray cavity ranges $L_{\rm kin} \sim 10^{42} - 10^{46}$ erg/s.
On the other hand, the radiative power of the radio jet is a few orders of magnitude smaller than the kinetic power \citep[e.g.,][]{Birzan_2004}.
The discrepancy between the radio and the kinetic powers of the lobes
indicates that the energy flux of the jet is carried by non-radiative particles, such as relativistic or non-relativistic protons, or electromagnetic fields.
The energy composition in the radio lobes is still under debate with the possibility that it depends on the morphological type of the jet, i.e., FR I or FR II.
Some studies suggest that the energy density of CRPs can be comparable to those of CREs and the magnetic field \citep[e.g.,][]{Hardcastle_2002,Godfrey_Shabala_2013,Croston_2018,Turner_2018}.
The maximal CRP injection rate (averaged over the Hubble time) can be estimated from a powerful radio jet:
\begin{equation}\label{eq:LCR_jet}
    L_{p}^{\rm jet} \lesssim 10^{42.5}~{\rm erg/s}\left(\frac{\chi_p}{0.3}\right)\left(\frac{L_{\rm kin}}{10^{45}~{\rm erg/s}}\right)\left(\frac{t_{\rm jet}/t_{\rm H}}{0.01}\right),
\end{equation}
where $\chi_p$ is the energy fraction of CRPs inside the lobe, and the ratio $t_{\rm jet}/t_{\rm H}$ is the duty cycle of the jet activity. 
In general, massive GCs host several radio galaxies, thus allowing $L_{\rm CR}^{\rm jet}$ to reach as high as $\sim10^{43}$ erg/s.
Note that some of the previous studies have assumed even larger injection rates from AGN jets, reaching up to $10^{45}$ erg/s \citep[e.g.,][]{Fang_Murase_2018}.
As the Lorentz factors of jets are typically 10--30 at launch, even cold gas in the jets can supply relativistic protons, i.e. CRPs, in the ICM. However, the adiabatic cooling during the jet evolution (deceleration and expansion into ICM) is an uncertain factor for the injection rate.
\par

\subsubsection{Accretion shocks}
Although we have focused on the internal sources in the central region of GCs, the cosmological accretion shocks can supply CRPs as well.
The accretion shock is considered to be formed around the cluster virial radius and is characterized by a high Mach number ($M_{\rm s}\sim10^{2}-10^3$).
Some studies have reported indications of CR acceleration at cluster virial shocks \citep[][]{Keshet_2017,Ilani_2024}.
Assuming that the CR acceleration efficiency is $\eta_{\rm acc}\sim1$ \%, one can estimate the CR injection luminosity as follows \citep[e.g.,][]{Keshet_2023}:
\begin{align}\label{eq:LCR_sh}
    L_{p}^{\rm shock} &\sim 10^{44}~{\rm erg/s}\nonumber \\
    & \times \left(\frac{\eta_{\rm acc}}{0.01}\right)
    \left(\frac{M_{\Delta_{\rm vir}}}{10^{15}~M_\odot}\right)^{\frac{5}{3}},
\end{align}
while observational evidence of the accretion shock is not significant at present \citep[e.g.][]{2022MNRAS.514.1645A,2023A&A...678A.197M}.
\par

\subsubsection{Discrepancy in CRP Injection Rate}

As we have discussed, considering various types of CRP sources, the CRP injection rate into ICM of $L_{p}\sim10^{43}-10^{44}$ erg/s is not optimistic too much.
However, our primary-electron reacceleration model, which can successfully explain the observed RH properties, allows $L_{p}\lesssim10^{40}-10^{41}$ erg/s.
The order-of-magnitude gap between the two values would be attributed to assumptions in our simulation in Sect.~\ref{sec:stat_RH}.
In the following sections, we demonstrate that $L_p^{\rm inj}$ exhibits a variation of only a factor of $\approx3$ across the models with different parameters. 


\subsection{Model C: No reacceleration at the pre-merger state}\label{sec:modelC}

\begin{figure*}
    \centering
    \plottwo{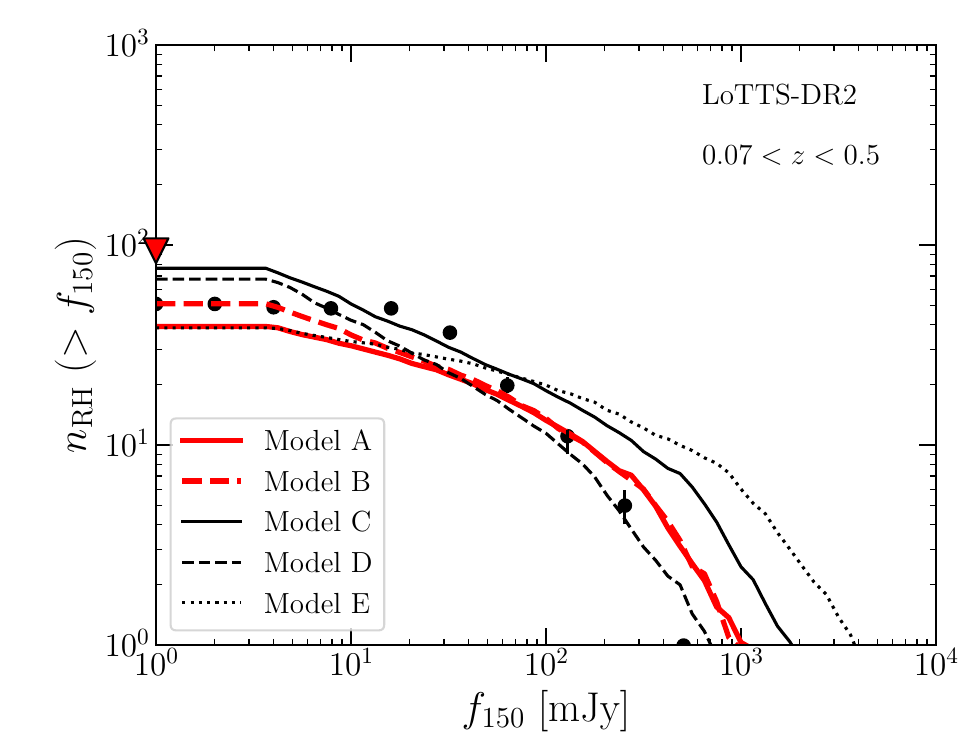}{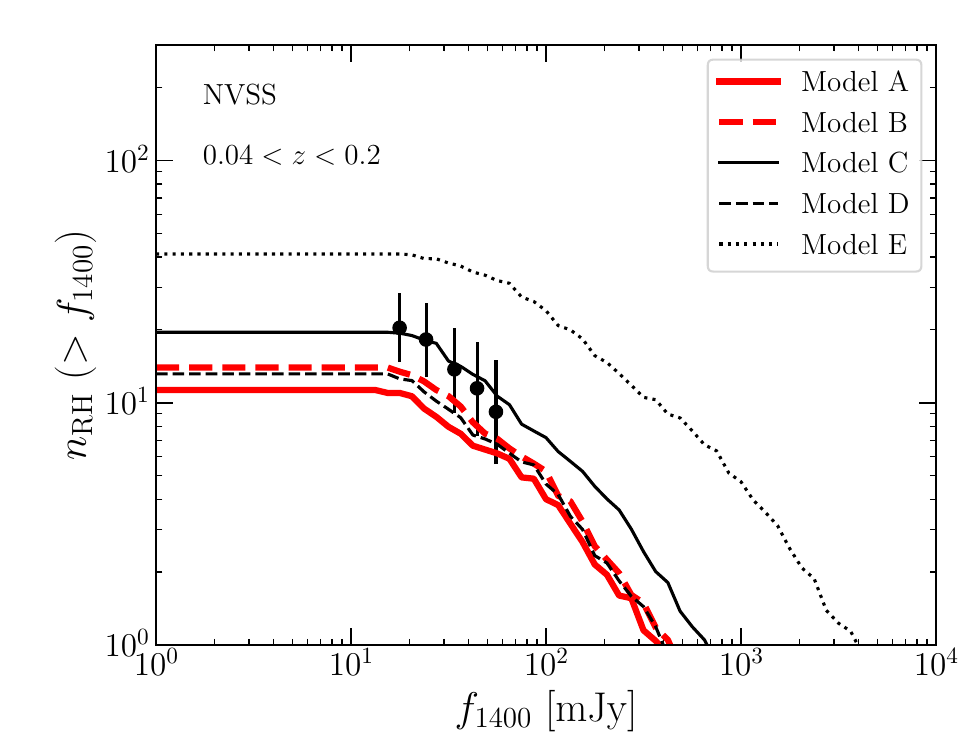}
    \caption{Left: same as Fig.~\ref{fig:nRH}, but for the simultaneous plot of five models A, B, C, D and E. Shown are the mean of the $N = 100$ trials of the sampling. The red thick lines show the results for models A and B.
    Right: RH number count in the NVSS survey at 1.4 GHz in the redshift range of $0.04 < z< 0.2$.}
    \label{fig:nRH_models}
\end{figure*}

Numerical simulations indicate the presence of ICM turbulence even in the pre-merger state \citep[e.g.,][]{Vazza_2011}.
However, the turbulent reacceleration in the pre-merger state can hardly be constrained by the RH observations.
In Sect.~\ref{sec:turb_prof}, we have assumed $f_{\rm comp}^{\rm rel} = 0.3$, which results in a gentle reacceleration by the TTD mechanism with an acceleration timescale of several Gyrs (Fig.~\ref{fig:turb}). 
While this weak reacceleration does not affect the CRE spectrum, the reacceleration of CRPs in the pre-merger state is not negligible due to the inefficiency of the cooling.
This may be the reason for the relatively low upper-limit of $L_{p}^{\rm inj}$ in our fiducial model (model B).
We test the case where the pre-merger reacceleration is completely negligible, $f_{\rm comp}^{\rm pre} = 0$.
This model is designated as Model C with its parameters except for $L^{\rm inj}_s$ and $f_{\rm comp}^{\rm pre}$ are the same as those of model B.
\par

With Model C, we obtain a rough upper-limit for the CRP injection rate, which is $\approx3$ times larger than that of Model B, i.e., $L_p^{\rm inj} = 1.5 \times 10^{41}$ erg/s.
In Fig.~\ref{fig:nRH_models}, we compare the RH number counts in various models. 
The right panel shows the expected number count adjusted for the NVSS survey at 1.4 GHz, where the sensitivity and the detection criterion are modeled in accordance with the methodology in \citet{paperII}.
Model C predicts a slightly larger number of RHs than Model B because of larger values for the injection rate.
The number count in Model C is compatible with the results of both LoTSS and NVSS observations.
A larger $L_p^{\rm inj}$ than that in model C causes the transition to the secondary-electron reacceleration model generating a larger number of RHs than observed numbers.
\par

The CRP spectrum and radial profile for the example clusters are presented in Fig.~\ref{fig:CRP}.
The peak of the CRP spectrum slightly shifts towards smaller $p$ compared to Model B, which can be attributed to the reduced effect of the reacceleration.
The difference in the median radial profile of $\epsilon_{\rm CRP}/\epsilon_{\rm th}$ between Model B and C is within a factor of $\approx2$.
These results indicate that the reacceleration in the pre-merger state is not the primary driver of the low injection rate in the reacceleration model (Sect.~\ref{sec:inj_problem}).

\subsection{Model D: Fast spatial diffusion of GeV CRs}\label{sec:modelD}

In the previous sections, we have assumed that spatial diffusion of CRs occurs due to the pitch-angle scattering in the presence of Alfv\'enic turbulence.
In this model, the diffusion coefficient of GeV CR is approximately $D_{rr} \sim 10^{29.5}$ cm$^2$/s in the cluster environment \citep[e.g.,][]{Murase_2013,Fang_Olinto_2016,paperI}.
The diffusion coefficient scales with the particle energy as $D_{rr} \propto p^{2-w}$, where $w$ is the index of the turbulent spectrum.
In the Kolmogorov turbulence, the particles with an energy below $\sim1$ PeV are unable to diffuse across the cluster volume within the Hubble time, resulting in their confinement within GCs.
This confinement effect due to the fiducial value of $D_{rr}$ can reduce the allowed upper-limit of the CRP injection rate.
\par

The mechanism of CR transport in the ICM is poorly understood, and a larger value for the diffusion coefficient may be possible in different models.
For example, the scattering of CRs associated with the turbulent dynamo and reconnection by the solenoidal turbulence at the Alfv\'en scale, at which the turbulent velocity becomes comparable to the Alfv\'en velocity, results in a spatial diffusion with \citep[][]{BL16}
\begin{equation}\label{eq:D_sol}
    D_{rr} \sim 10^{31}~{\rm cm^2/s}\left(\frac{\psi}{0.5}\right)\left(\frac{l_{\rm A}}{1~{\rm kpc}}\right),
\end{equation}
where $\psi$ is the mean free path normalized by the Alfv\'en scale $l_{\rm A}$ (typically 0.1-1 kpc).
Otherwise, the turbulent motion of the ICM directly transports CRs coupled with the fluid.
In this case, the effective diffusion coefficient becomes $D\sim v_{\rm A}l_{\rm A}\sim 10^{28}-10^{30}$ cm$^2$/s \citep[e.g.,][]{Wiener_2013}.
\par

In Model D, we explore the case of rapid spatial diffusion of GeV CRs. We adopt a constant diffusion coefficient of $D_{rr} = 10^{31}$ cm$^2$/s. 
For simplicity, we neglect the possible parameter dependence of $D_{rr}$ on redshift, mass, and the dynamical state (pre-merger or post-merger). 
In this model, the diffusion coefficient for CRs with $E \gtrsim 30$ TeV is {\it smaller} than that predicted by the pitch-angle scattering model.
The remaining model parameters, except for $L_p^{\rm inj}$, are the same as those in Model B.
As illustrated in Fig.~\ref{fig:nRH_models}, Model D (black dashed line) can accept three times greater $L_p^{\rm inj}$ than that in model B consistent with the observed number count.
Similarly to Model C, the diffusion coefficient does not significantly impact the results.
The upper-limit of the CR injection rate remains below the previous expectations (Sect.~\ref{sec:inj_problem}).
\par

A comparison of CRP energy spectrum for various models is presented in Fig.~\ref{fig:CRP}.
The CRP energy spectrum in Model D is harder than that of Model B.
This is due to the reduction of the diffusive escape of higher-energy CRPs with $E_p \gtrsim 30$ TeV.
The CRP energy density is enhanced due to the larger number of $E_p \gtrsim $ PeV CRPs.
Even with the higher CRP energy density, the harder CRP spectrum reduces the effective number of CRPs responsible for the secondary-electron injection, so that the predicted RH number can be consistent with the observations.
The peak of $\epsilon_{\rm CRP}/\epsilon_{\rm th}$ is shifted towards smaller radii, as the seeping out of CRPs to the cluster outskirts is reduced. The saturation of the turbulent reacceleration occurs in the radial range of $0.2<r/R_{200}<0.5$, where $\epsilon_{\rm CRP}$ becomes comparable to $\epsilon_{\rm turb}$.

\subsection{Model E: Small minimum momentum of primary CREs}\label{sec:modelE}

Concerning the minimum momentum of the primary CREs, our models A and B assume $p_e^{\rm min}/(m_ec) = 300$. 
The cooling timescale of CREs, i.e., the lifetime of CREs, naturally produces a CRE spectrum with a maximum around $p_e^{\rm min}/(m_ec) \approx 300$ in the ICM \citep[e.g.,][]{Sarazin_1999,Brunetti_Jones_review}.
However, it may be possible that $p_e^{\rm min}/(m_ec) < 300$ CREs are provided in the ICM without suffering from the severe cooling at the vicinity of the source.
In this section, we test Model E, in which $p_e^{\rm min}/(m_ec) = 1$. 
We use the same spectral index as Model A, $\alpha_{\rm CR} = 2.2$, and consider the case of $L_p^{\rm inj} = 0$ erg/s.
\par

Assuming $L_{e}^{\rm inj} = 1.5\times10^{41}$ erg/s for the CRE injection rate in Model E, we obtain the total number of RHs consistent with the LoTSS-DR2 survey.
However, as seen from Fig.~\ref{fig:nRH_models}, this model overpredicts the number of RHs with large radio fluxes ($f_{150} > 10^3$ mJy).
In this model, the spectrum of the seed CREs before the reacceleration extends below $p/(m_{e}c) = 300$, so there are larger amounts of low-energy seed CREs compared to Model A.
During the reacceleration phase, the CREs are accelerated to the momentum of $p/(m_{e}c) \sim 10^4$.
The increase in the radio luminosity during the reacceleration is more pronounced in Model E than in Model A due to the larger amount of seed CREs.
If one adopts a smaller $L_e^{\rm inj}$ in order to reduce the number of too bright RHs, the model underpredicts the total number of RHs.
In order to achieve a better fit to the observation with the $p/(m_{e}c) = 1$ model, a fine-tuning of the reacceleration parameters would be necessary.
\par

\subsection{Possible solution to the low CRP injection problem}

We have shown that the injection of only the primary CREs with $L_e^{\rm inj}=10^{40}$--$10^{41}$ erg/s, like model A, is significant to reproduce the RH statistics. Considering the low ratio of CREs to CRPs in our galaxy and the cooling effect before escape into the ICM, this relatively low value of the {\it CRE} injection rate can be acceptable. However, the allowed upper limit of the {\it CRP} injection rate, $L_p^{\rm inj} \lesssim L_e^{\rm inj}$, is enigmatic as we have discussed in this section.

\subsubsection{Significant energy loss at the source?} \label{sec:CRPloss}
The injection rates presented in Sect~\ref{sec:inj_problem} may be overestimated since we neglect the energy loss of CRPs at their source environments.
Concerning the CR injection from starburst galaxies (SBGs) (Eq.~(\ref{eq:LCR_Fe})), one of the most significant energy-loss processes is the hadronic $pp$ collision.
Due to the substantial gas column density, the SBGs are expected to be nearly calorimetric for CRPs, i.e., CRPs protons lose their energies via the $pp$ collision before escaping via diffusion or advection \citep[e.g.,][]{Pohl_1994,Paglione_1996}.
Indeed, the {\it Fermi} satellite has detected GeV gamma rays from nearby SBGs, including M82, NGC 253, and Arp 220 \citep[e.g.,][]{Abdo_2010,Fermi_2012_starforming,Peng_2016,Ajello_2020}.
The fluxes are well explained with the hadronic model with a calorimetric fraction of $f_{\rm cal}\sim 0.3-1$ \citep[e.g.,][]{Lacki_2011,Wang_2018,Krumholz_2020,Ambrosone_2024}.
\par

Furthermore, CRs should suffer from the adiabatic loss in the galactic wind when escaping from SBGs.
The galactic wind is a promising mechanism for the chemical enrichment of the surrounding ICM \citep[e.g.,][]{Elbaz_1995,Fujita_2008,Mernier_Biffi_2022}.
The wind launched from the central region of SBGs ($r \sim 1$ kpc) with a power of $E_{\rm w} \sim 10^{60}$ erg \citep[e.g.,][]{Thompson_2024} can expand to the size of $r\sim 100$ kpc under the typical thermal pressure of the ICM ($P_{\rm ICM}\sim 5\times10^{-3}~{\rm keV}$/cm$^3$).
If the CRs are tightly coupled to the wind gas, the CR energy density is reduced by a factor of $\sim10^{-6}$ due to the adiabatic loss.
Due to these fatal energy losses, the tension between the limit (Tab.~\ref{tab:models}) and the expectation for $L_{p}$ (Eq.~(\ref{eq:LCR_Fe})) can be alleviated.
Note that some studies have suggested that the termination shock of the galactic wind can (re)accelerate the CRs up to very-high energies \citep[e.g.,][]{Jokipii_1987,Merten_2018,Peretti_2022}.
A consist model for the chemical enrichment and the CR injection due to the starburst-driven galactic wind should be developed in future studies.
\par
\subsubsection{Small filling factor of turbulent region?}\label{sec:patchy}
According to recent MHD simulations, the distribution of turbulence in the ICM is inhomogeneous. The turbulent kinetic energy flux spans more than six orders of magnitude within a 1 Mpc$^3$ box region (see Fig.~2 of \citet{megahalo}). 
As a consequence of the greater efficiency of reacceleration in regions with a larger turbulent flux, the CRs in these regions contribute the majority of the observed emission.
In addition, the magnetic field can be amplified in such regions due to the turbulent dynamo.
In \citet{megahalo}, we find that the most of the observed flux of the mega halo of Abell 2255 can be produced by the reaccelerated CREs in highly turbulent parts whose volume-filling factor is $\sim5$\%. 
The contribution from the remaining $\sim$95 \% of the region is almost negligible. 
The small effective volume of the re-acceleration zone allows a higher injection rate of CRs compared to our homogeneous (for a given $r$) model.
This may also be a solution to the problem of a low CRP injection rate in the reacceleration model.
However, it should be noted that the above discussion is largely dependent on the assumption of particle diffusion, which was neglected in \citet{megahalo} for simplicity.
The limit on $L_{p}^{\rm inj}$ should be revisited taking into account the effects of both the patchy distribution of turbulence and the CR diffusion.

\section{Conclusions}\label{sec:conclusion}
The formation of GCs is accompanied by the CR production by SNRs or AGN activities in the member galaxies.
CRPs provided by these accelerators are preserved in the hot and dilute ICM over a cosmological timescale.
The observation of multi-messenger non-thermal emissions is crucial for studying the co-evolution of galaxies and GCs imprinted in the CRP distribution.
The turbulent reacceleration of CRs is a promising mechanism for explaining various observational properties of RHs.
However, it is still ambiguous to what extent the secondary electrons produced by the $pp$ collisions of CRPs contribute to the radio emission.
In this paper, we succeed in producing models of the CR evolution in the ICM with turbulent reacceleration consistently with the cosmological structure formation, star formation history, galaxy number density profile in GCs, RH statistics, radio observations of the Coma cluster, background neutrino upper limit given by IceCube, and hydrodynamical simulations of the ICM turbulence. Based on those models, we examine the constraints on the CRP injection rate. 
\par



The major update from our previous method \citep{paperI,paperII,paperIII} is to solve the FP equations for CREs and CRPs (Eq.~\ref{eq:FP_mass}) considering the evolution of cluster mass and the ICM profiles.
We build the 1D spherically-symmetric model of the ICM, using the observed profiles of the thermal pressure, temperature, and magnetic field.
The method allows for following the CR injection history during the cosmological evolution of GCs.
Emissions from a variety of GCs with different mass evolution histories are simulated.
We assume that the primary CRs are injected from internal sources, such as AGNs and SBGs. 
The injection rate is proportional to the cluster mass and the SFR density.
\par

The main results of the fiducial reacceleration models (models A and B) are summarized as follows: 
\begin{itemize}
    \item The conventional primary-electron reacceleration model is compatible with the observed spectrum and brightness profile of the Coma RH, assuming that Coma is in the early stage of the reacceleration process.
    \item We recover the occurrence and the mass-radio power relation in a population of RHs with a threshold mass ratio of $\xi_{\rm th} = 0.2$ for triggering the reacceleration and the reacceleration duration of $T_{\rm dur} = 4t_{\rm eddy}\approx 3$Gyr.
    \item The CRP injection luminosity, $L_{p}^{\rm inj}$, up to $\sim3\times10^{41}$ erg/s is consistent with the spectrum of Coma. This upper limit is significantly lower than the previous estimates. This discrepancy is attributable to the updates introduced for the sake of a realistic modeling, including an extended injection duration and a narrower injection profile, as well as the reacceleration in the pre-merger state.
    \item The statistical properties of RHs also provide a tight constraint on $L_{p}^{\rm inj}$. For the required duration $T_{\rm dur} \approx 3$Gyr, to avoid too large and too long injection of secondary electrons, the injection rate of CRPs is required to be as small as $L_{p}^{\rm inj} \lesssim 10^{41}$ erg/s. The upper limit is a few orders of magnitude smaller than the rate expected from star formation in member galaxies or the AGN activity.
        A shorter $T_{\rm dur}$ can be possible (see Appendix \ref{app:T_dur}), but the required CR injection rate is not largely enhanced.
   \item Neglecting properties of individual RHs, we roughly reproduce the observed data, although there are some mismatches to the data points (Figs.~\ref{fig:Coma} and \ref{fig:nRH}). Even when accounting for the uncertainty in the model details, the upper limit of $L_{p}^{\rm inj}$ may not exceed $10^{42}$ erg/s in the primary CRE model.
    \item Due to the inefficiency of the radiative cooling, CRPs are efficiently accelerated up to $E_p \sim $PeV by the hard-sphere reacceleration. The resulting CRP spectrum is very hard, as presented in Fig.~\ref{fig:CRP}. 
    \item In Model B, the expected intensity of the neutrino background is consistent with the upper limit by \citet{IceCube_2022_stack}, so the nearby massive GCs cannot be the dominant source of the neutrino background. The neutrino limit excludes $L_{p}^{\rm inj} > 10^{42}$ erg/s in the reacceleration model.
    \item We do not rule out the possibility of the secondary-electron reacceleration model, where the primary electrons are negligible. In this model, the optimal values for the reacceleration parameters would be significantly different from those used in this study \citep[][]{paperII}. A reexamination of this model is desired in future research. 
\end{itemize}

The low CR injection rate obtained in our model may depend on the various assumptions about CR injection spectrum, diffusion, and reacceleration, which are however poorly constrained.
In Sect.~\ref{sec:discussion}, we discuss the parameter dependence by testing three models C, D, and E. 
The results are summarized as follows:
\begin{itemize}
    \item Model C; we conclude that the reacceleration in the pre-merger state is not the major factor for the low injection rate. The CRP injection rate can be $L_{p}^{\rm inj} \approx 1.5\times10^{41}$ erg/s when the pre-merger reacceleration is completely negligible.
    \item Model D; we test the case with an energy independent and large spatial diffusion coefficient of $D_{rr} = 10^{31}$ cm$^2$/s. The limiting $L_{p}^{\rm inj}$ is not different from that in Model B very much. 
    \item Model E; the minimum momentum of primary CREs  is reduced as $p_{0,e}/(m_ec)  = 1$ from 300 in model A. This model predicts an increased number of RHs with high fluxes, which appears to be inconsistent with the observed data.
\end{itemize}


Throughout this study, we demonstrate that the theoretical modeling of RHs can provide a meaningful constraint on the CRPs in GCs.
The apparent gap in the CRP density between the CR reservoir model \citep[e.g.,][]{Volk_1996} and the reacceleration model imposes revisions to those models.
The injection rate of CRPs in the reservoir model may be overestimated due to the neglect of the fatal energy loss of CRPs ``at the source", which occurs due to adiabatic expansion or hadronic interaction.
As for the reacceleration model, we may have overlooked important physics, such as the patchy distribution of the turbulent region seen in the MHD simulations \citep[e.g.,][]{BV20}.
Future studies including the inhomogeneous turbulence or more sophisticated wave-particle interaction theory with significant statistics of multiwavelength and multi-messenger observations will provide unique insights into the co-evolution of galaxies and GCs and the mechanism of particle acceleration in the ICM.

\begin{acknowledgments}
We would like to thank Yutaka Fujita for helpful discussions.
This work is supported by the joint research program of the Institute for Cosmic Ray Research (ICRR), the University of Tokyo, and KAKENHI grant Nos. JP23KJ0486 (K.N.), 22K03684, 23H04899, and 24H00025 (K.A.).
\end{acknowledgments}

\appendix

\section{RH statistics with a short $T_{\rm dur}$}\label{app:T_dur}
In Sect.~\ref{sec:stat_RH}, we show that our models with the reacceleration duration of $T_{\rm dur} = 4t_{\rm dur} \approx 3$Gyr and the CR injection rate listed in Tab.~\ref{tab:models} are consistent with the RH number count in the LoTSS-DR2 survey.
In this section, we discuss the dependence of the CR injection rate on the parameter $T_{\rm dur}$, examining the case of $T_{\rm dur} = 2t_{\rm eddy} \approx 1.5{\rm Gyr}$. 
For simplicity, we only consider the case without CRPs and secondaries.
In short, we find that the total RH number can be reproduced with $L_e^{\rm inj} \approx 1.5\times10^{41}$ erg/s, i.e., 5 times larger than that in Model A. However, this 
provides a less optimal fit to the RH flux distribution.

\par

In this work, we use two parameters, $\xi_{\rm th}$ and $T_{\rm dur}$, which control the occurrence of RHs in the stochastic merger tree (Sect.~\ref{sec:stat_RH}).
In general, a shorter $T_{\rm dur}$ leads to a smaller number of detectable RHs as long as $T_{\rm dur}$ is longer than the radiative cooling time of $\sim$GeV CREs ($t_{\rm cool} \approx$ 500 Myr).
For a shorter $T_{\rm dur}$, a smaller $\xi_{\rm th}$ (low threshold for reacceleration) is required to explain the observed number of RHs.
In this section, we adopt the set of $(T_{\rm dur},\xi_{\rm th}) = (2t_{\rm eddy}, 0.1)$, which can roughly explain the observed RH fraction as shown by \citet{Cassano_2016}.
It is worth noting that $\xi_{\rm th}$ should be as small as 0.01 when $T_{\rm dur} \approx 500$ Myr in the primary-electron reacceleration model, as reported in our previous work \citep[][]{paperII}.
However, it may be unlikely that a minor merger with $\xi\sim0.01$ results in a giant halo with an extension comparable to that of the major progenitor cluster.
We explore the parameter space in between that model and Model A in Sect.~\ref{sec:stat_RH}.

\par

\begin{figure*}
    \centering
    \plottwo{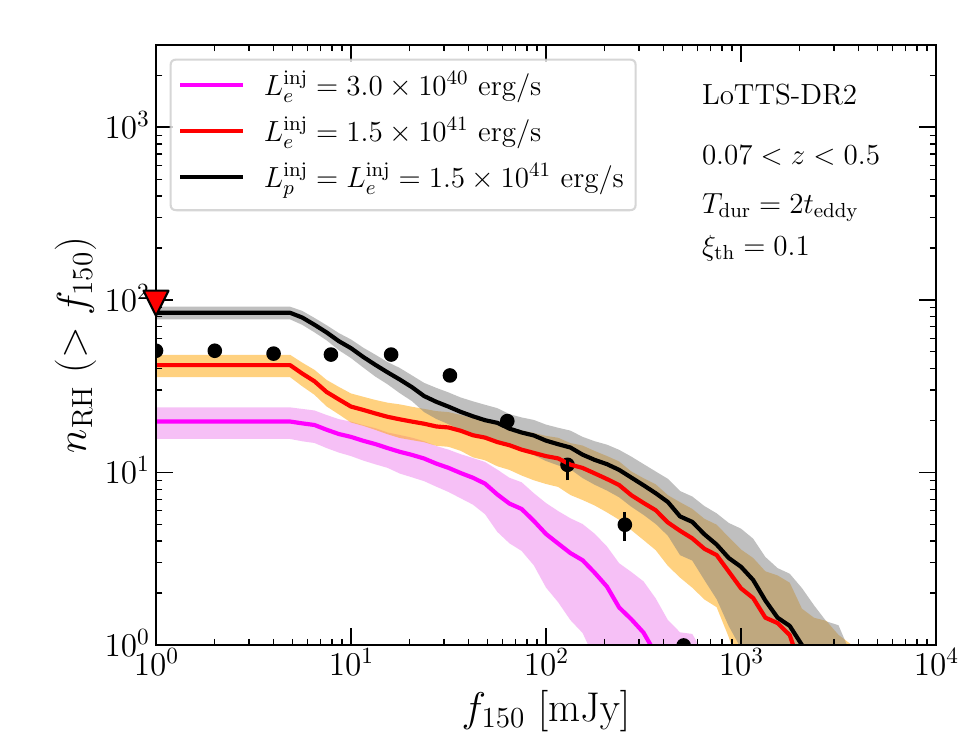}{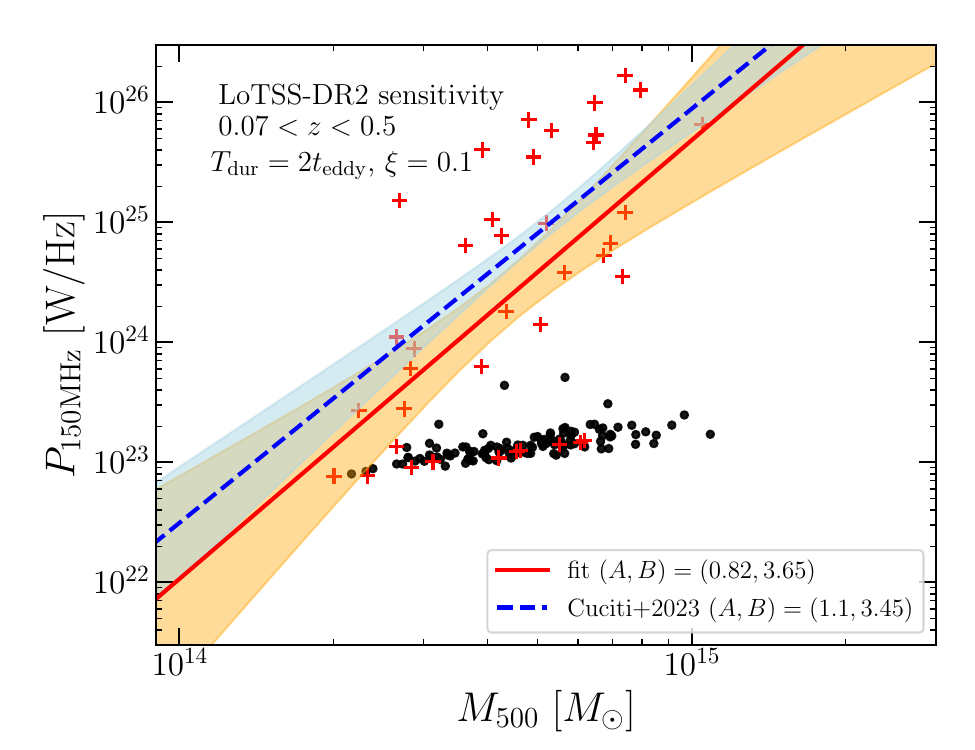}
    \caption{
     Left: same as Fig.~\ref{fig:nRH}, but for the model with $T_{\rm dur} = 2t_{\rm eddy}$ and $\xi = 0.1$. The red and magenta lines represent the cases of $L_e^{\rm inj} = 1.5\times10^{41}$ erg/s and $L_e^{\rm inj} = 3.0\times10^{40}$ erg/s, respectively,  without CRP injection. The model including CRPs ($L_p^{\rm inj} = L_e^{\rm inj}  = 1.5\times10^{41}$ erg/s) is shown with the black line and gray area. Right: same as Fig.~\ref{fig:P-M}, but for $T_{\rm dur} = 2t_{\rm eddy}$, $\xi = 0.1$, and $L_e^{\rm inj} = 1.5\times10^{41}$ erg/s. The red and black points represent RHs and non-RHs in our mock sample. The blue line shows the best-fit relation for the LoTSS-DR2 RHs \citep[][]{Cuciti_2023}.}
    \label{fig:App}
\end{figure*}

In Fig.~\ref{fig:App}, we show the flux distribution of LoTSS-detectable RHs in the current model.
Assuming the same injection rate as Model A, i.e., $L_{e}^{\rm inj}\approx 3\times10^{40}$ erg/s, the model underpredicts the number of RHs (magenta line and area).
Instead, $L_{e}^{\rm inj}\approx 1.5\times10^{41}$ erg/s is required.
For a shorter $T_{\rm dur}$ 
than that in Model A, the probability of detecting RHs induced by $\xi \geq 0.2$ mergers becomes smaller, and those induced by $0.1<\xi<0.2$, for which the reacceleration efficiency is low, contribute to the RH statistics mainly.
Such a model requires a larger CR injection rate.

\par

Although the total number of RHs is consistent with the observation in this case, the match to the observed flux distribution is worse than Model A (Fig.~\ref{fig:App} left panel).
We can find a bump in the low flux region ($f_{\rm 150 MHz}< 10$ mJy).
Those RHs are mostly located at low redshift and exhibit relatively small radio powers. 
The presence of these RHs is evident in the radio power–mass diagram (right panel).
The radio powers of all clusters in our sample are elevated due to the increase of $L_e^{\rm inj}$, and the figure shows that some of the nearby relaxed clusters exhibit the detectable emission and are classified as RHs (red points).
Thus, the distribution of $P_{\rm 150MHz}$ of RHs in the diagram is more dispersed than that of Model A (Fig.~\ref{fig:P-M}) and there is a significant overlap between RHs and non-RHs.
However, this might be in conflict with the claims made by \citet{Cuciti_2023} that there is a distinct separation in $P_{\rm 150MHz}$ between RHs and non-RHs, and that the bimodality is not merely caused by the sensitivity limit but rather an intrinsic phenomenon.
On the other hand, one should note the recent discoveries of diffuse radio emission enveloping the classical mini-halos in relaxed clusters \citep[e.g.,][]{Bonafede_2014,Venturi_2017,Savini_2019,Biava_2021,Knowles_2022,Bruno_2023,Lusetti_2024,vanWeeren_2024}.
A quantitative comparison of the $P_{\rm 150MHz}$ distribution between observation and our model is interesting, but it is beyond the scope of this work.

\par

As in Model A and B, the presence of the secondary CREs affects the RH statistics when $L_p^{\rm inj}/L_e^{\rm inj} \gtrsim 1$ (the black line and gray region in Fig.~\ref{fig:App}). 
We again obtain the order-of-magnitude constraint on the CRP injection rate as $L_p^{\rm inj}\lesssim 10^{41}$ erg/s.
Thus, the limit on $L_p^{\rm inj}$ in the short $T_{\rm dur}$ model is not significantly different from that in the original model in Sect.~\ref{sec:stat_RH}.



\bibliography{D}{}
\bibliographystyle{aasjournal}

\end{document}